\documentclass[10pt,a4paper]{article}
\usepackage[T1]{fontenc}
\usepackage{microtype}
\usepackage{amsmath,amssymb}
\usepackage{amsfonts}
\usepackage{mathrsfs}
\usepackage[table]{xcolor}  
\usepackage{booktabs}
\usepackage{array}
\usepackage{multirow}
\usepackage{longtable}
\usepackage{tabularx}
\usepackage{xltabular}
\usepackage{graphicx}
\usepackage[export]{adjustbox}
\usepackage{pdflscape}
\usepackage{rotating}
\usepackage{caption}
\usepackage{subcaption}
\usepackage{authblk}
\usepackage{etoolbox}
\usepackage{epstopdf}
\usepackage{orcidlink}
\usepackage{textcomp}
\usepackage[margin=1.5cm,top=1.5cm]{geometry}
\usepackage{hyperref}
\usepackage[numbers,sort&compress]{natbib}
\hypersetup{colorlinks=true, linkcolor=blue, citecolor=blue, urlcolor=blue}
\providecommand{\keywords}[1]{%
  \par\vspace{0.5ex}%
  \noindent\textbf{Keywords: }#1\par
}

\numberwithin{equation}{section}
\numberwithin{figure}{section}
\numberwithin{table}{section}
\newcolumntype{C}{>{\centering\arraybackslash}X}
\newcolumntype{L}{>{\raggedright\arraybackslash}X}
\newcolumntype{R}{>{\raggedleft\arraybackslash}X}
\renewcommand{\arraystretch}{1.3}
\begin{document}

\begin{center}
\LARGE{Gravitational Wave Signatures of Periodic Orbits around a Schwarzschild-like Black Holes Submerged in an Exponential Density Dark Matter Profile \\\;}
\par\end{center}

\begin{center}
{\bf Mohammad Reza Alipour\orcidlink{0000-0001-8074-7865}}\footnote{\bf mohamad.alipour.1994@gmail.com; mr.alipour@stu.umz.ac.ir}\\
{\it School of Physics, Damghan University, P.~O.~Box 3671641167, Damghan, Iran}\\
{\it Center for Theoretical Physics, Khazar University, 41 Mehseti Street, Baku, AZ1096, Azerbaijan}
\end{center}

\begin{center}
{\bf Saeed Noori Gashti\orcidlink{0000-0001-7844-2640}}\footnote{\bf sn.gashti@du.ac.ir; saeed.noorigashti70@gmail.com}\\
{\it School of Physics, Damghan University, P.~O.~Box 3671641167, Damghan, Iran}
\end{center}

\begin{center}
{\bf Mohammad Ali S. Afshar\orcidlink{0009-0001-3133-5992}}\footnote{\bf m.a.s.afshar@gmail.com}\\
{\it Department of Physics, Faculty of Basic Sciences, University of Mazandaran P. O. Box 47416-95447, Babolsar, Iran}\\
{\it School of Physics, Damghan University, P.~O.~Box 3671641167, Damghan, Iran}\\
{\it Center for Theoretical Physics, Khazar University, 41 Mehseti Street, Baku, AZ1096, Azerbaijan}
\end{center}

\begin{center}
{\bf Behnam Pourhassan\orcidlink{0000-0003-1338-7083}}\footnote{\bf b.pourhassan@du.ac.ir}\\
{\it School of Physics, Damghan University, P.~O.~Box 3671641167, Damghan, Iran}\\
{\it Center for Theoretical Physics, Khazar University, 41 Mehseti Street, Baku, AZ1096, Azerbaijan}
\end{center}

\begin{abstract}
We study a Schwarzschild-like black hole embedded in an exponential-sphere (ESM) dark matter halo, characterized by a halo mass $M_0$ and a scale radius $r_0$. We first show that the halo's effect on the marginally bound orbit (MBO) and innermost stable circular orbit (ISCO) is richer than a simple shift: the characteristic radii and energies vary non-monotonically with $r_0$, dipping below their Schwarzschild values before recovering, while the angular momenta decrease monotonically; as a function of $M_0$, the response can even reverse sign depending on how extended the halo is, with only the ISCO energy remaining monotonic throughout. We classify periodic orbits by the rational frequency ratio $q=\omega_\phi/\omega_r-1=w+v/z$ and find that $r_0$ and $M_0$ leave clearly distinguishable imprints on the orbit spectrum. Using the numerical kludge framework, we compute the corresponding extreme-mass-ratio-inspiral (EMRI) waveforms and show that varying $r_0$ produces a strong, monotonic effect enlarging the orbits, lengthening the radial period, and introducing a clear dephasing while comparable variations in $M_0$ leave the signal nearly unchanged unless $M_0$ becomes a sizable fraction of the black hole mass.  Together, these results indicate that EMRI waveforms can, in principle, disentangle the total mass of a dark matter halo from its spatial extent, offering a strong-field probe of the dark matter distribution around supermassive black holes.
\end{abstract}


\tableofcontents

\section{Introduction}
It was Einstein's general relativity that first explained gravity not as a force, but as a consequence of spacetime curvature~\cite{Einstein1916}. This fresh perspective fundamentally reshaped our understanding of the universe. Among its most striking predictions was the existence of black holes objects that are not only the final stage of massive stars but also one of the theory's most challenging theoretical outcomes~\cite{Schwarzschild1916,Kerr1963}. In recent years, two major breakthroughs have tested the validity of general relativity across different domains: the first was the direct detection of gravitational waves from merging compact binaries~\cite{AbbottGW150914,AbbottGW151226}, and the second was the horizon-scale imaging of two supermassive objects at the centres of M87 and the Milky Way~\cite{EHTM87,EHTSgrA}. These successes have shown that general relativity performs well in both weak and strongly gravitational conditions. Today, black holes serve as natural laboratories for examining alternative theories of gravity and for exploring the boundaries of general relativity in astrophysical settings. For this reason, rigorous strong-field tests remain vital for understanding gravitational-wave physics, horizon structure, and the very nature of gravity itself~\cite{Will2014,Psaltis2020}.

In this context, gravitational-wave analysis provides a remarkably powerful tool. Future space-based interferometers such as LISA, Taiji, and TianQin~\cite{AmaroSeoaneLISA,HuWuTaiji2017,GongTianQin2021} will be sensitive to a wide range of low-frequency sources. Among the most important of these are extreme mass-ratio inspirals (EMRIs), where a stellar-mass compact object spirals around a supermassive black hole~\cite{Hughes2001,AmaroSeoaneEMRI,Babak2017}. These systems emit long-lived, low-frequency gravitational waves whose waveforms carry detailed information about the particle dynamics and the surrounding spacetime geometry, including any departures from circular motion. Any modification to the background spacetime alters the orbital frequencies and resonance conditions; these changes, in turn, affect the stability of circular orbits, the structure of zoom-whirl dynamics, and ultimately the observed EMRI signal~\cite{GlampedakisKennefick2002,RuangsriHughes2014}. Hence, the waveforms from such systems offer a direct window into horizon structure and gravitational nature, acting as sensitive probes that leave their imprints throughout the long inspiral phase~\cite{BarackEtAl2019}.

A substantial part of the gravitational waveforms produced by these systems arises from \emph{periodic orbits}. These bound trajectories are fundamental to EMRI dynamics and provide an effective way to describe the long inspiral phase through a sequence of periodic configurations~\cite{GrossmanLevin2009,MisraLevin2010}. A periodic orbit occurs when a massive particle returns to its starting point after a finite number of radial ($\omega_{r}$) and azimuthal ($\omega_{\phi}$) oscillations, exhibiting the characteristic zoom-whirl morphology and satisfying an $r$--$\phi$ resonance condition. Levin and Perez-Giz characterised each such orbit by the rational frequency ratio $\omega_{\phi}/\omega_{r}$ and three integers $(z,w,v)$ the zoom, whirl, and vertex numbers~\cite{LevinPerezGiz2008,Levin2009}. This classification offers a systematic framework for modelling periodic orbits, analysing their zoom-whirl dynamics, resonance structures, and spectral features along the inspiral. Although this approach was originally developed for Schwarzschild and Kerr spacetimes using the topological integers $(z,w,v)$~\cite{LevinPerezGiz2008,Levin2009,BambhaniyaEtAl2021,RanaMangalam2019}, it has since been extended to many other backgrounds. For example, the gravitational waveforms of periodic orbits in EMRIs have been computed for quantum-corrected and regular geometries~\cite{ChenYang2025,LiKuang2026,HuaEtAl2026}, for the $\gamma$-metric~\cite{ZhangZhuGamma2026}, for Einstein--\AE ther and Kalb--Ramond black holes~\cite{LuZhu2025,JuniorEtAl2025}, for black holes free of a Cauchy horizon~\cite{WangCauchy2025}, for non-commutative Schwarzschild black holes with a Lorentzian matter distribution~\cite{HeidariEtAl2026}, for Schwarzschild-like black holes embedded in Dehnen-type and Hernquist dark-matter halos~\cite{ShokirovEtAl2026,HeidariAraujoFilho2026Hernquist,AlloqulovDehnen2025,HaroonZhu2025}, for a Schwarzschild--Bertotti--Robinson black hole immersed in a uniform magnetic field~\cite{XamidovEtAl2026}, for a non-commutative-inspired black hole surrounded by quintessence~\cite{AhmedQuintessence2026}, and for spinning test particles in Schwarzschild spacetime~\cite{UktamovEtAl2026}. In all these studies, the numerical kludge scheme~\cite{BabakKludge2007,PoissonWill2014} has been adopted, which combines a numerically integrated geodesic trajectory with the quadrupole formula to construct the corresponding EMRI signals. Two general conclusions can be drawn from this body of work. First, modifications to the background geometry are faithfully reflected in the emitted radiation, so that EMRI waveforms genuinely probe the underlying spacetime. Second, a clear dichotomy emerges between two categories of modification: environmental influences, such as a surrounding dark-matter halo, tend to expand the allowed region of bound motion and enlarge the periodic orbits, whereas corrections that are intrinsic to the compact object such as charge, non-commutativity, or short-distance quantum effects act in the opposite direction, shrinking the orbits and lowering their characteristic energies and angular momenta. Against this background, the short-distance quantum-gravity modification considered in the present work finds a natural place.

The work presented in \cite{5001} examines timelike geodesic motion, periodic orbits, and the associated gravitational-wave signatures around a Bonanno-Reuter regular black hole, which emerges within the framework of asymptotically safe gravity. In this setting, the scale dependence of Newton's coupling replaces the classical central curvature singularity with a regular de Sitter core. A key finding is that the entire strong-field dynamics is governed by a single dimensionless combination, $\alpha/M^{2}$. As this parameter increases, both the marginally bound orbit and the innermost stable circular orbit shift toward smaller radii, while their corresponding angular momenta decrease, along with the ISCO energy. The allowed region in the $(E,L)$ plane correspondingly moves toward lower values, favouring more tightly bound configurations. These results clarify how the asymptotically safe correction modifies the orbital dynamics, energy, and angular momentum of timelike test particles near the black hole. The study also investigates the gravitational-wave signals generated by a stellar-mass compact object moving along periodic orbits around a supermassive Bonanno--Reuter black hole. Using the numerical kludge approach, the orbital trajectories and the corresponding gravitational-wave polarizations are calculated. It is found that increasing $\alpha/M^{2}$ contracts the periodic orbits, shortens the orbital periods, and slightly reduces the waveform amplitudes. This sensitivity depends on the orbital topology and grows with the complexity parameter $(z,w)$. The resulting spectra predominantly lie in the millihertz frequency band, with several characteristic-strain peaks exceeding the sensitivity curves of future space-based detectors such as LISA, Taiji, and TianQin. These findings suggest that short-distance quantum-gravity modifications of Newton's coupling may leave observable imprints on extreme mass-ratio inspiral gravitational-wave signals.

The structure of this paper proceeds as follows. In Section~\ref{sec:metric}, we introduce the Schwarzschild-like black hole solution surrounded by an exponential density dark matter halo. Section~\ref{sec:mbo-isco} derives the equations of motion for timelike test particles, constructs the corresponding effective potential, and determines the characteristic MBO and ISCO radii as functions of the dark matter model parameters. In Section~\ref{sec:periodic}, we turn to periodic orbits; these are classified using the rational number $q$, and their zoom-whirl morphology is analysed in detail, with particular attention to how the dark matter environment modifies the resonance conditions. Section~\ref{sec:waveforms} presents the numerical kludge waveforms we have constructed, discusses their detectability by future space-based interferometers, and compares the resulting gravitational wave signatures with those from other dark matter profiles and strong-field modifications. Throughout this work, we adopt geometrized Planck units, setting $c=k_{B}=G_{N}=\hbar=1$, and use the mostly-plus signature for the metric.
\section{The Model}
\label{sec:metric}
To derive tractable field equations, we restrict ourselves to a static, spherically symmetric geometry and adopt the usual diagonal form for the line element. In coordinates $(t, r, \theta, \phi)$, we write \cite{5000}
\begin{equation}
ds^{2} = -A(r) dt^{2} + \frac{dr^{2}}{B(r)} + r^{2} \left( d\theta^{2} + \sin^{2}\theta \, d\phi^{2} \right),
\label{eq:metric}
\end{equation}
with the metric functions satisfying
\begin{equation}
A(r) = B(r) = 1 - \frac{2m(r)}{r},
\label{eq:metricfunctions}
\end{equation}
where $m(r)$ denotes the mass function.
Our goal is to solve the Einstein field equations and obtain a spherically symmetric black hole metric embedded within the ESM dark matter halo. For the pure halo spacetime, the field equations take the form
\begin{equation}
R^{\mu}_{\nu} - \frac{1}{2} \delta^{\mu}_{\nu} R = \kappa T^{\mu}_{\nu},
\label{eq:efe}
\end{equation}
where $T^{\mu}_{\nu} = \mathrm{diag}[-\rho, p_{r}, p_{t}, p_{t}]$ is the energy-momentum tensor of the ESM profile, which is generally anisotropic. For the static, spherically symmetric ansatz (\ref{eq:metric}), the independent Einstein equations reduce to the compact set
\begin{align}
\kappa T^{t}_{t} &= A \left( \frac{1}{r^{2}} + \frac{1}{r} \frac{A'}{A} \right) - \frac{1}{r^{2}}, \label{eq:tt}\\
\kappa T^{r}_{r} &= A \left( \frac{1}{r^{2}} + \frac{1}{r} \frac{A'}{A} \right) - \frac{1}{r^{2}}, \label{eq:rr}\\
\kappa T^{\theta}_{\theta} &= \frac{1}{2}A \left[ \frac{A'' A}{A^{2}} + \frac{2}{r} \left( \frac{A'}{A} \right) \right]. \label{eq:thetatheta}
\end{align}
Upon substituting the metric functions and the energy-momentum tensor components into the field equations, we arrive at the simple relations
\begin{align}
\rho &= -p_{r} = \frac{m'(r)}{4\pi r^{2}}, \label{eq:rho}\\
p_{t} &= -\frac{m''(r)}{8\pi r}. \label{eq:pt}
\end{align}
In this work, we employ the ESM to model the density distribution of dark matter halos. This is a simple phenomenological profile that has proven effective in describing low-surface-brightness and dwarf galaxies. We take the energy density to be
\begin{equation}
\rho(r) = \rho_{0} \, e^{-r/r_{0}},
\label{eq:density}
\end{equation}
where $\rho_{0}$ and $r_{0}$ represent the characteristic density and scale radius of the halo, respectively.
The mass profile for the combined system -- a black hole plus dark matter -- is then given by
\begin{equation}
m(r) = \int_{M_{\mathrm{BH}}}^{r} 4\pi \tilde{r}^{2} \rho(\tilde{r}) \, d\tilde{r} = M_{\mathrm{BH}} + M_{0} \left[ 1 - e^{-r/r_{0}} \left( 1 + \frac{r}{r_{0}} + \frac{r^{2}}{2r_{0}^{2}} \right) \right],
\label{eq:massprofile}
\end{equation}
where we have defined
\begin{equation}
M_{0} \equiv 8\pi \rho_{0} r_{0}^{3}.
\label{eq:M0}
\end{equation}
Here $M_{0}$ denotes the total mass of the dark matter distribution (the halo mass). In the asymptotic limit $r \to \infty$, the mass function approaches the sum of the black hole and halo masses, $m \to M_{\mathrm{BH}} + M_{0}$. Near the centre, as $r \to 0$, we recover $m \to M_{\mathrm{BH}}$.
Correspondingly, the dressed black hole metric takes a Schwarzschild-like form
\begin{equation}
A(r) = 1 - \frac{2M_{\mathrm{BH}}}{r} - \frac{2M_{0}}{r} \left[ 1 - e^{-r/r_{0}} \left( 1 + \frac{r}{r_{0}} + \frac{r^{2}}{2r_{0}^{2}} \right) \right].
\label{eq:Afunction}
\end{equation}
The metric (\ref{eq:Afunction}) thus describes a one-parameter family -- or more precisely, a two-parameter family in terms of $\rho_{0}$ and $r_{0}$ -- of Schwarzschild-like black holes ``dressed'' by the ESM halo. This solution satisfies several useful consistency checks that confirm its physical viability. First, in the limit where the characteristic density vanishes, $\rho_{0} \to 0$, the dark matter halo is absent and the metric reduces exactly to the Schwarzschild geometry. Second, when the central black hole mass is set to zero, $M_{\mathrm{BH}} \to 0$, the solution recovers a regular pure halo spacetime without any central singularity. Third, after imposing the standard normalisation $A(\infty) = 1$, which ensures asymptotic flatness at spatial infinity, the coefficient of the $1/r$ term in the expansion of the metric function is unambiguously identified with the ADM mass $M = M_{\mathrm{BH}} + M_{0}$ of the full spacetime. These limiting cases provide strong evidence that the dressed metric (\ref{eq:Afunction}) correctly interpolates between the Schwarzschild geometry and the pure dark matter halo spacetime, and that the total mass parameter appearing in the asymptotic expansion has the physically expected interpretation as the sum of the black hole and halo contributions.
\section{Timelike Geodesics}
\label{sec:mbo-isco}
\begin{figure}[htbp]
\centering
\begin{subfigure}[b]{0.48\linewidth}
\centering
\includegraphics[width=\linewidth]{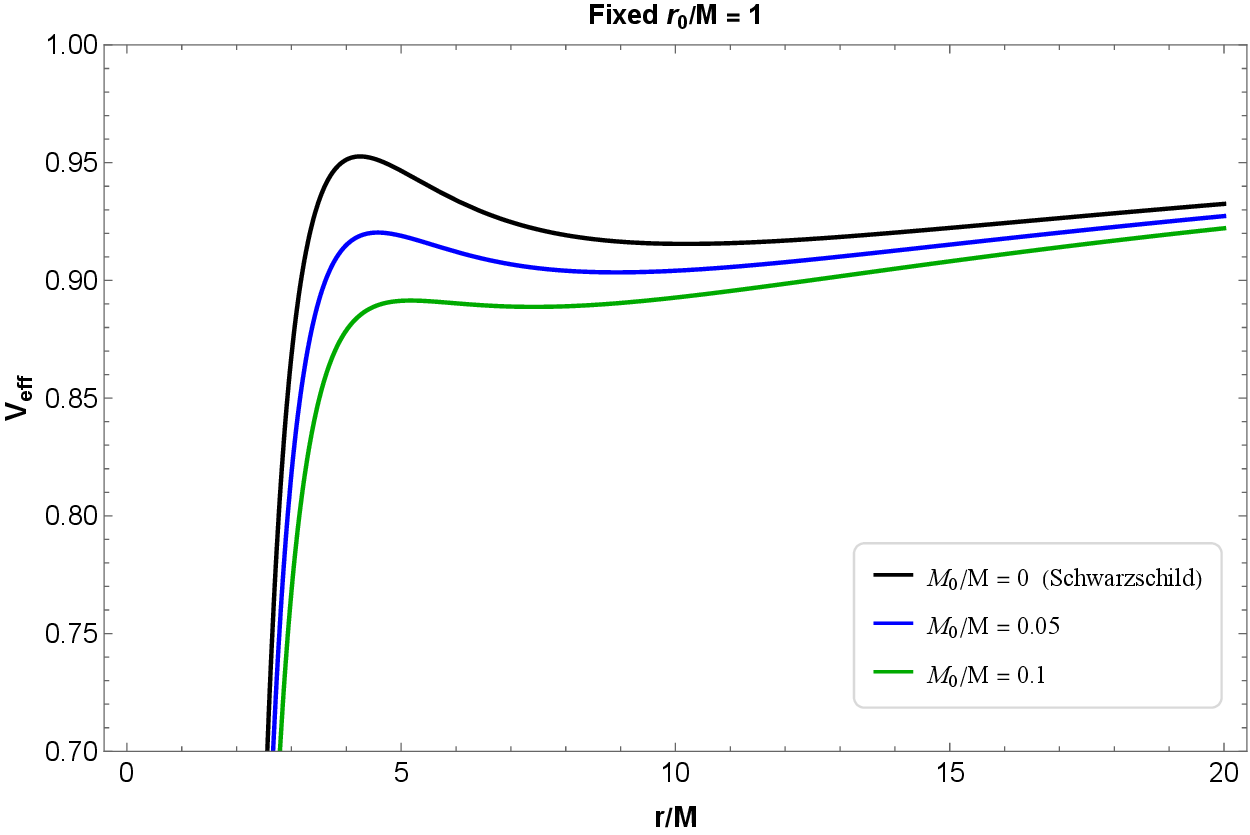}
\caption{}
\label{fig:Veff-M0}
\end{subfigure}
\hfill
\begin{subfigure}[b]{0.48\linewidth}
\centering
\includegraphics[width=\linewidth]{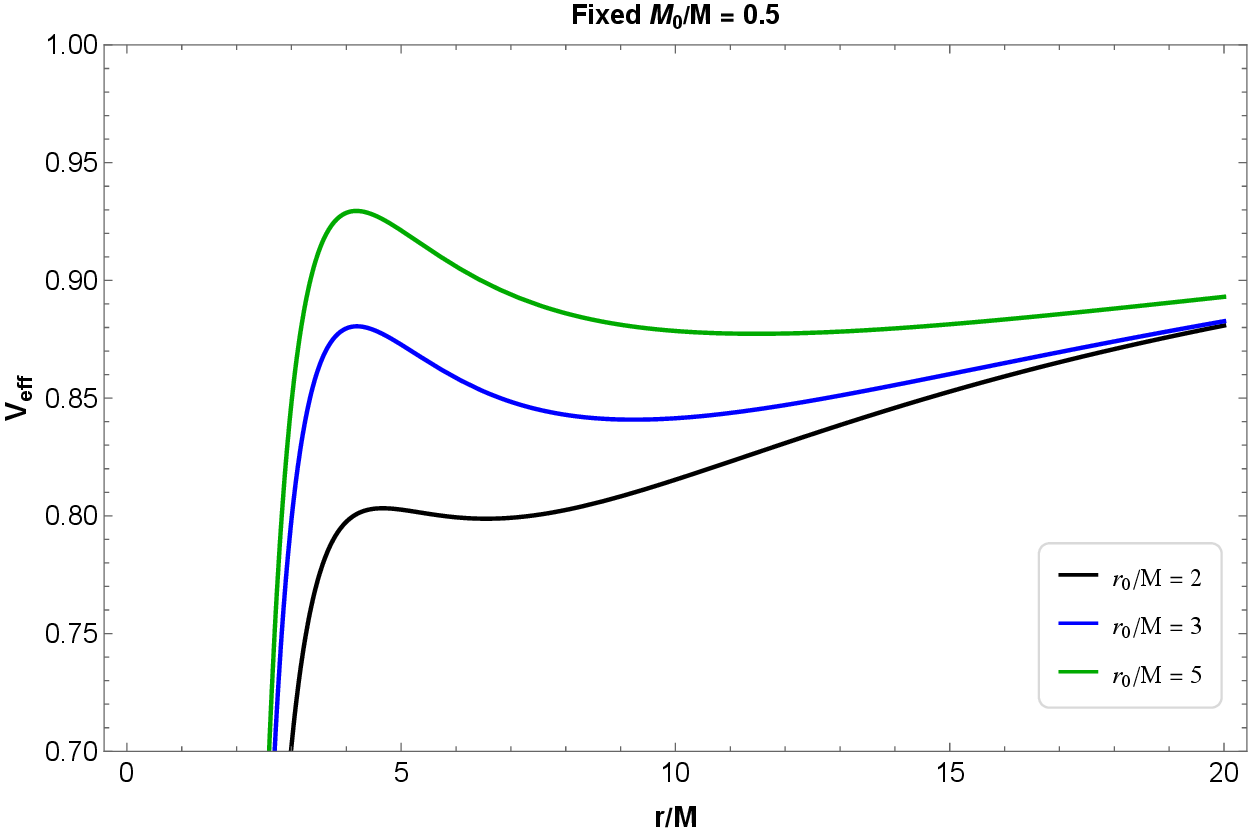}
\caption{}
\label{fig:Veff-r0}
\end{subfigure}
\caption{Effective potential $V_{\rm eff}(r)$ for timelike geodesics around the Schwarzschild-like black hole embedded in an exponential-sphere dark matter halo. Left panel: fixed halo scale radius $r_0/M=1$, for $M_0/M=0$ (Schwarzschild), $0.05$, and $0.1$. Right panel: fixed halo mass $M_0/M=0.5$, for $r_0/M=2$, $3$, and $5$.}
\label{fig:Veff-r0-M0}
\end{figure}

For a massive test particle moving in the equatorial plane ($\theta = \pi/2$), the geodesic Lagrangian takes the form
\begin{equation}
\mathcal{L} = \frac{1}{2} g_{\mu\nu} \dot{x}^{\mu} \dot{x}^{\nu} = \frac{1}{2} \left( -A(r) \dot{t}^{2} + \frac{\dot{r}^{2}}{A(r)} + r^{2} \dot{\phi}^{2} \right),
\label{eq:Lagrangian}
\end{equation}
from which the conserved specific energy $E = A(r) \dot{t}$ and angular momentum $L = r^{2} \dot{\phi}$ are derived. Imposing the timelike normalization condition $g_{\mu\nu} \dot{x}^{\mu} \dot{x}^{\nu} = -1$ reduces the radial dynamics to
\begin{equation}
\dot{r}^{2} = E^{2} - V_{\mathrm{eff}}(r),
\label{eq:radial}
\end{equation}
with the effective potential given by
\begin{equation}
V_{\mathrm{eff}}(r) = A(r) \left( 1 + \frac{L^{2}}{r^{2}} \right) = \bigg(1 - \frac{2M_{\mathrm{BH}}}{r} - \frac{2M_{0}}{r} \left[ 1 - e^{-r/r_{0}} \left( 1 + \frac{r}{r_{0}} + \frac{r^{2}}{2r_{0}^{2}} \right) \right] \bigg) \left( 1 + \frac{L^{2}}{r^{2}} \right).
\label{eq:Veff}
\end{equation}
Since the mass function $m(r)$ given in Eq.~\eqref{mass_ESM} is smooth and bounded for all $r>0$, the effective potential $V_{\mathrm{eff}}(r)$ is likewise smooth everywhere outside the horizon, $r>r_{h}$. The strong-field dynamics is governed by the two independent dimensionless ratios $M_{0}/M$ and $r_{0}/M$, which set, respectively, the total mass and the spatial extent of the dark matter halo.
As shown in Fig.~\ref{fig:Veff-r0-M0}, the two halo parameters act in opposite directions on the effective potential: increasing $r_0$ raises the height of the potential barrier, while increasing $M_0$ lowers it.

The domain of bound motion is delimited by the marginally bound orbit (MBO) and the innermost stable circular orbit (ISCO). For static, spherically symmetric metrics satisfying $g_{tt} g_{rr} = -1$, the circular orbit condition $\partial_{r} V_{\mathrm{eff}} = 0$ yields
\begin{equation}
L^{2}(r) = \frac{r^{3} A'(r)}{2 A(r) - r A'(r)}.
\label{eq:Lsquared}
\end{equation}
Substituting this expression back into the effective potential reduces the MBO and ISCO conditions to single equations in $r$. The MBO, defined by $V_{\mathrm{eff}}(r_{\mathrm{MBO}}) = 1$ and $\partial_{r} V_{\mathrm{eff}} = 0$, gives
\begin{align}
r_{\mathrm{MBO}} &= \frac{2 A(r_{\mathrm{MBO}}) [1 - A(r_{\mathrm{MBO}})]}{A'(r_{\mathrm{MBO}})}, \label{eq:MBO_r}\\
L_{\mathrm{MBO}} &= r_{\mathrm{MBO}} \sqrt{\frac{1 - A(r_{\mathrm{MBO}})}{A(r_{\mathrm{MBO}})}}. \label{eq:MBO_L}
\end{align}

\begin{figure}[htbp]
\centering
\begin{subfigure}[b]{0.48\linewidth}
\centering
\includegraphics[width=\linewidth]{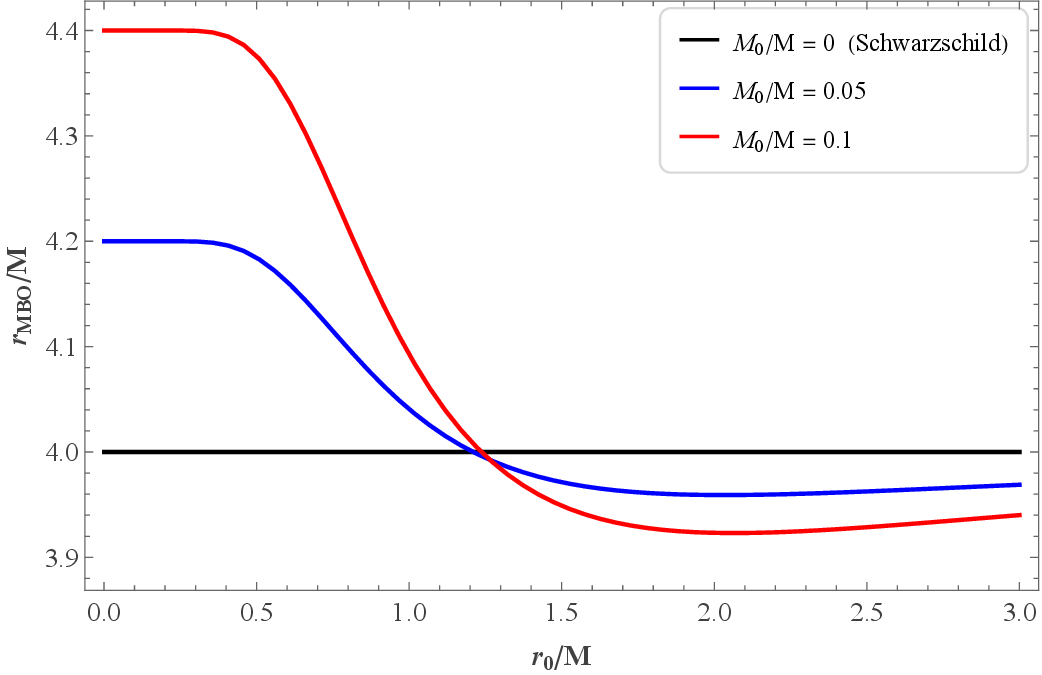}
\caption{}
\label{fig:rMBO-M0}
\end{subfigure}
\hfill
\begin{subfigure}[b]{0.48\linewidth}
\centering
\includegraphics[width=\linewidth]{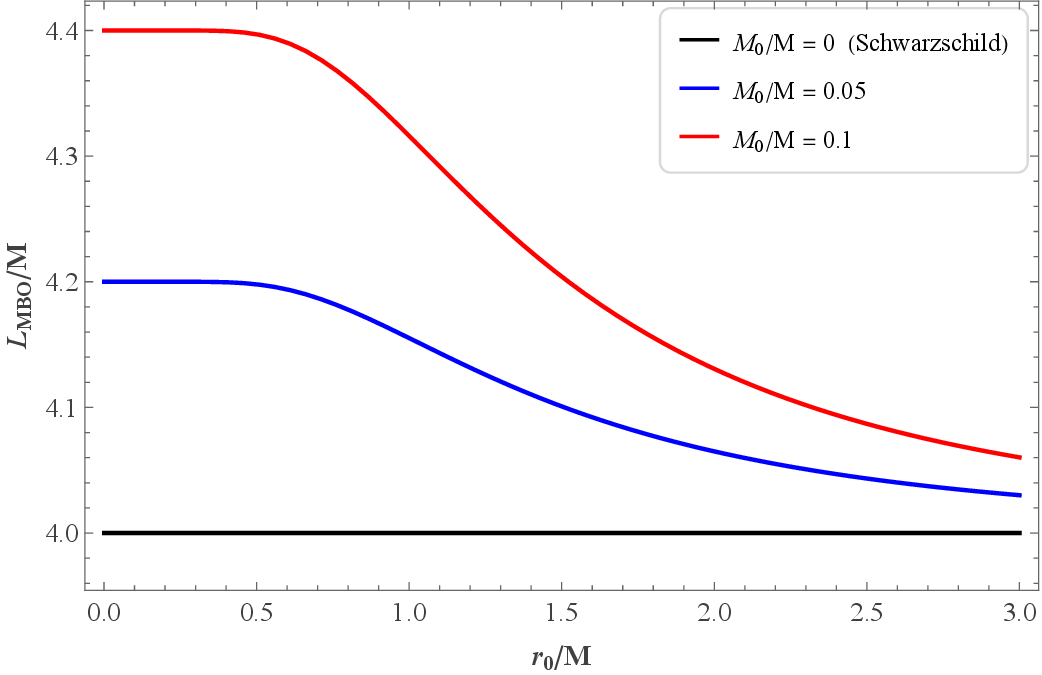}
\caption{}
\label{fig:LMBO-M0}
\end{subfigure}
\hfill
\begin{subfigure}[b]{0.48\linewidth}
\centering
\includegraphics[width=\linewidth]{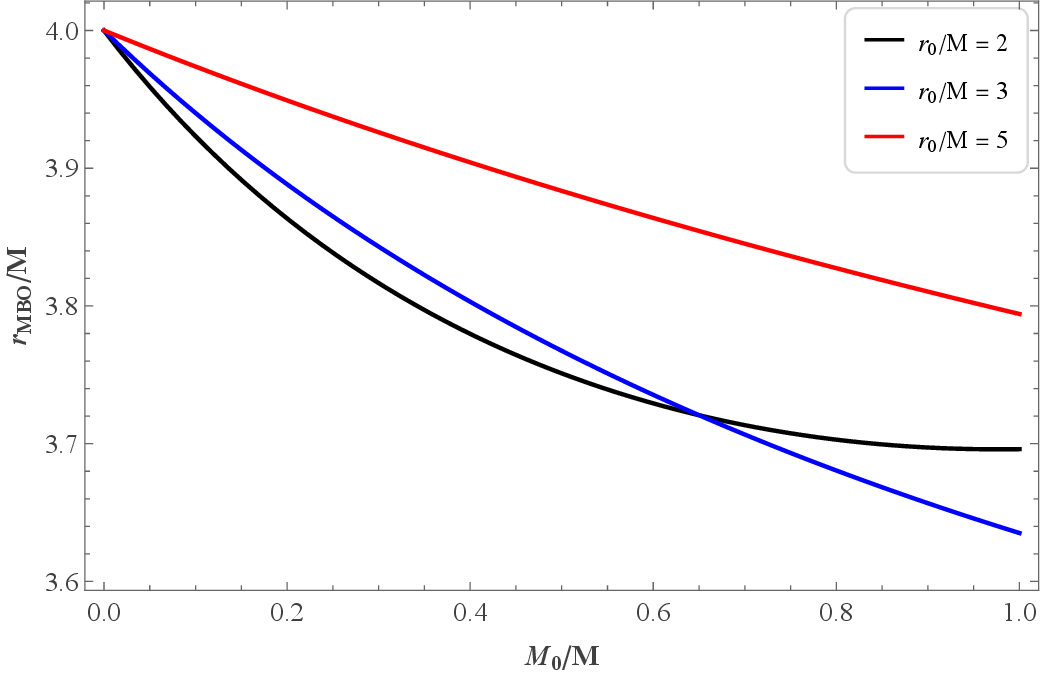}
\caption{}
\label{fig:rMBO-r0}
\end{subfigure}
\hfill
\begin{subfigure}[b]{0.48\linewidth}
\centering
\includegraphics[width=\linewidth]{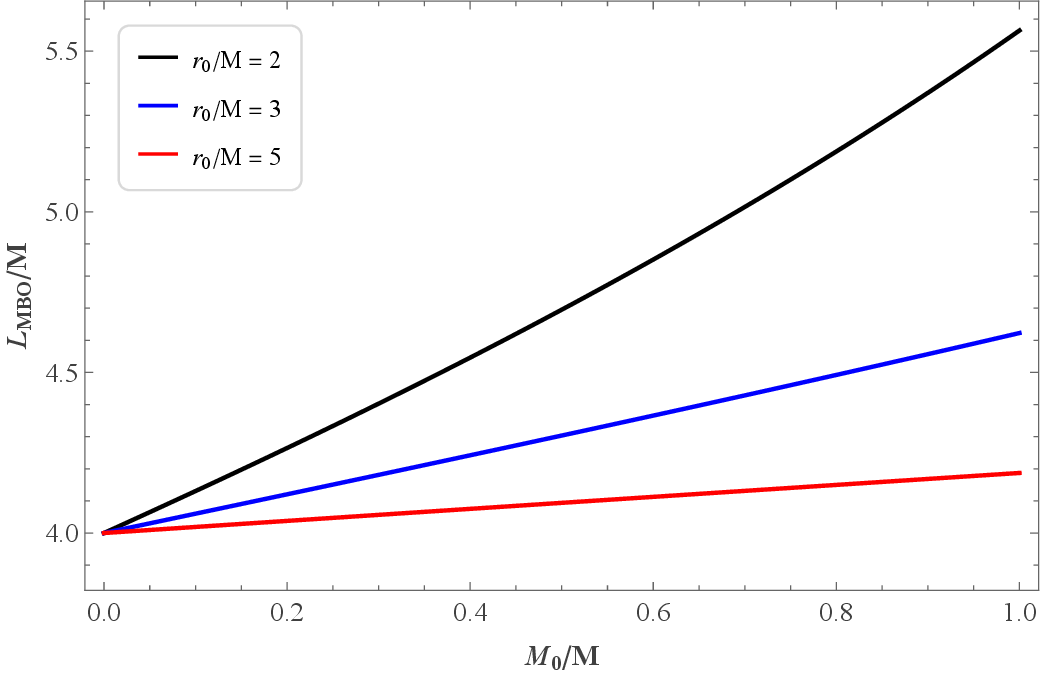}
\caption{}
\label{fig:LMBO-r0}
\end{subfigure}
\caption{Radius and angular momentum of the marginally bound orbit (MBO) for the Schwarzschild-like black hole embedded in an exponential-sphere dark matter halo. Top row: $r_{\rm MBO}/M$ (a) and $L_{\rm MBO}/M$ (b) versus $r_0/M$, for $M_0/M=0$ (Schwarzschild), $0.05$, and $0.1$. Bottom row: $r_{\rm MBO}/M$ (c) and $L_{\rm MBO}/M$ (d) versus $M_0/M$, for $r_0/M=2$, $3$, and $5$.}
\label{fig:MBO-vs-r0-M0}
\end{figure}
Figure~\ref{fig:MBO-vs-r0-M0} shows how the halo parameters affect the MBO. As a function of $r_0/M$ [panels (a),(b)], $r_{\rm MBO}$ is non-monotonic: it dips below the Schwarzschild value around $r_0/M\sim1.5$--$2$ before recovering, while $L_{\rm MBO}$ decreases monotonically toward the Schwarzschild limit with no such dip. As a function of $M_0/M$ [panels (c),(d)], both quantities vary monotonically, but their sensitivity to $M_0$ depends on $r_0$: a smaller $r_0$ gives a steeper response, with the $r_0/M=2$ and $3$ curves crossing near $M_0/M\sim0.65$, while the more extended halo ($r_0/M=5$) remains comparatively flat. This contrast indicates that the halo mass and its spatial extent leave distinguishable, rather than interchangeable, imprints on the MBO.

The ISCO requires the inflection condition $\partial_{r}^{2} V_{\mathrm{eff}} = 0$, which, together with the circular orbit condition, yields
\begin{equation}
r A(r) A''(r) - 2 r [A'(r)]^{2} + 3 A(r) A'(r) = 0.
\label{eq:ISCO_condition}
\end{equation}
The root $r_{\mathrm{ISCO}}$ of this equation determines the characteristic angular momentum and energy:
\begin{align}
L_{\mathrm{ISCO}} &= \sqrt{\frac{r_{\mathrm{ISCO}}^{3} A'(r_{\mathrm{ISCO}})}{2 A(r_{\mathrm{ISCO}}) - r_{\mathrm{ISCO}} A'(r_{\mathrm{ISCO}})}}, \label{eq:LISCO}\\
E_{\mathrm{ISCO}} &= \sqrt{\frac{2 A(r_{\mathrm{ISCO}})^{2}}{2 A(r_{\mathrm{ISCO}}) - r_{\mathrm{ISCO}} A'(r_{\mathrm{ISCO}})}}. \label{eq:EISCO}
\end{align}

\begin{figure}[h!]
  \centering
  \begin{subfigure}[b]{0.32\textwidth}
    \centering
    \includegraphics[width=\linewidth, height=\linewidth]{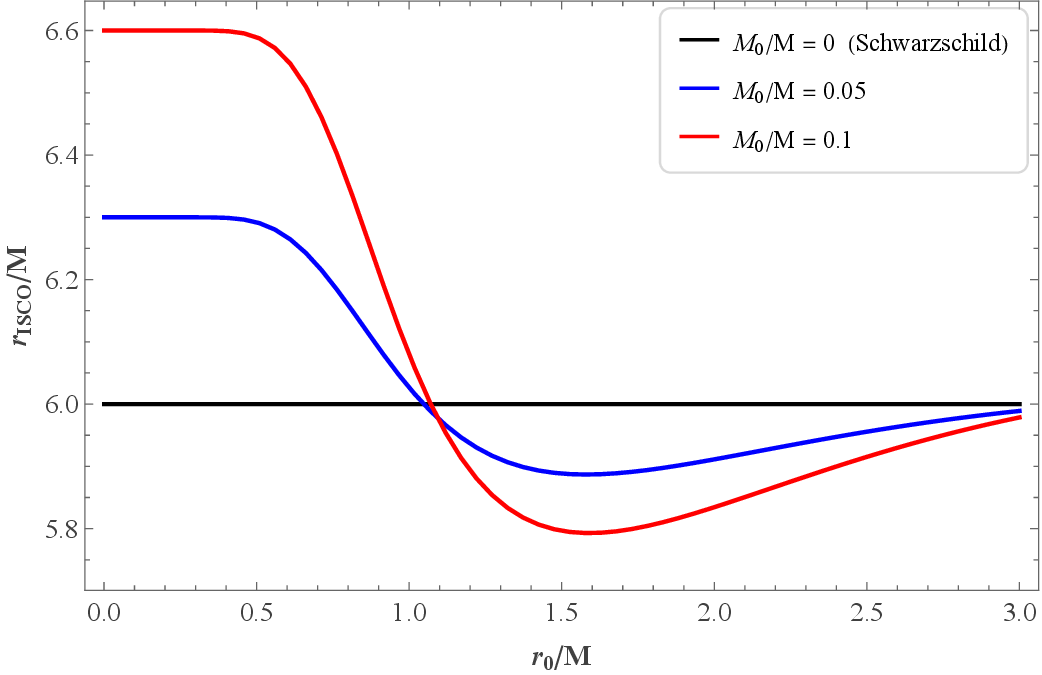}
    \caption{}
    \label{fig:rISCO-M0}
  \end{subfigure}
  \hfill
  \begin{subfigure}[b]{0.32\textwidth}
    \centering
    \includegraphics[width=\linewidth, height=\linewidth]{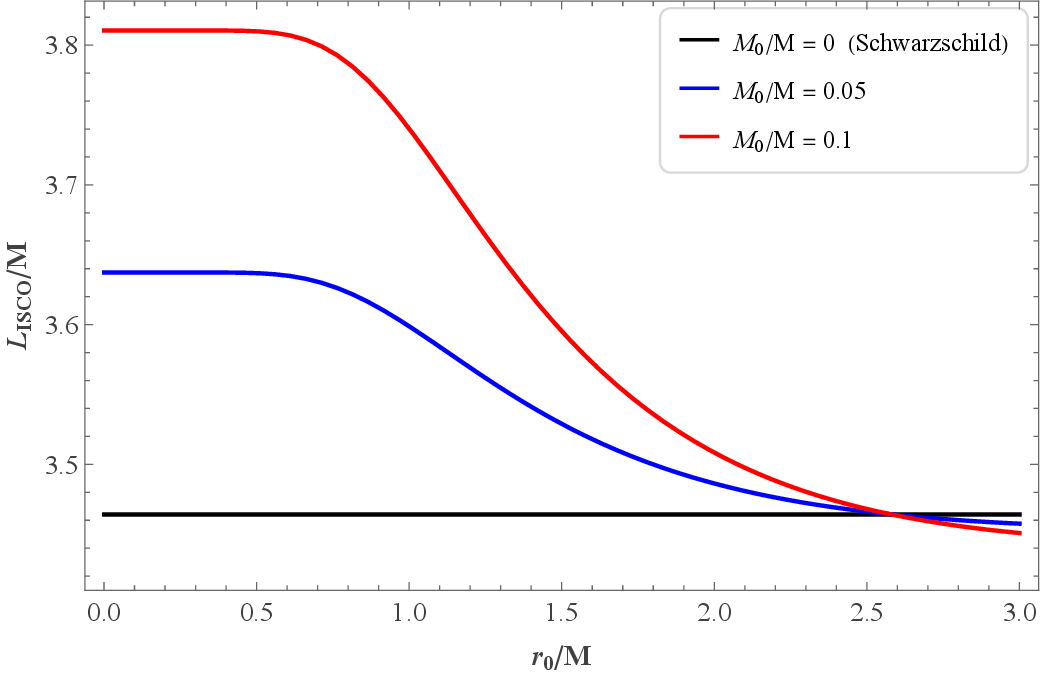}
    \caption{}
    \label{fig:LISCO-M0}
  \end{subfigure}
  \hfill
  \begin{subfigure}[b]{0.32\textwidth}
    \centering
    \includegraphics[width=\linewidth, height=\linewidth]{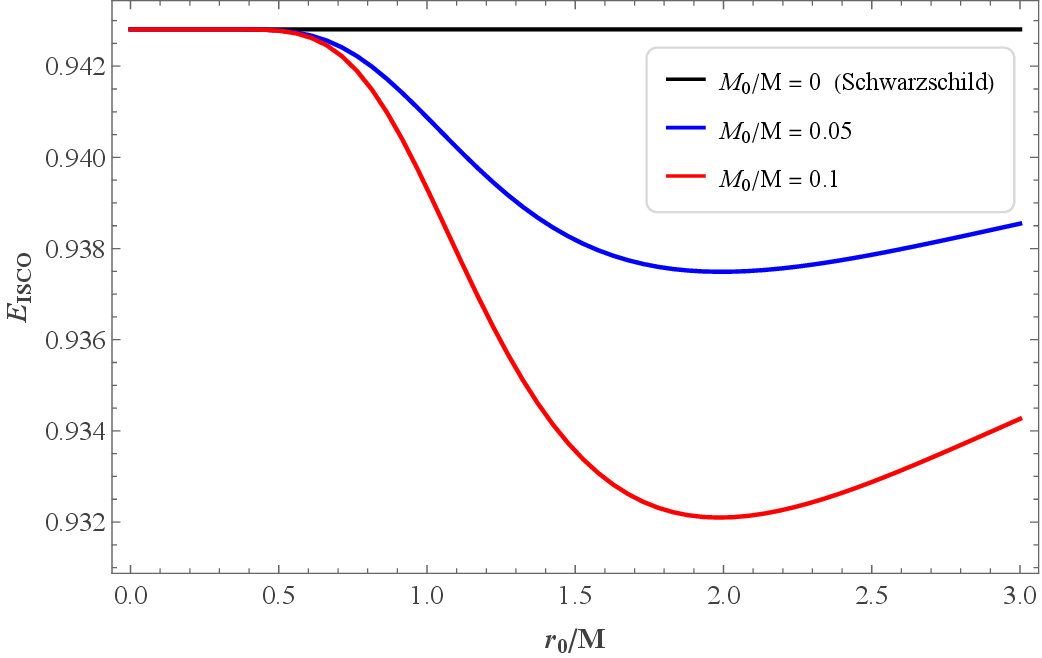}
    \caption{}
    \label{fig:EISCO-M0}
  \end{subfigure}

  \vspace{-0.6cm}

  \begin{subfigure}[b]{0.32\textwidth}
    \centering
    \includegraphics[width=\linewidth, height=\linewidth]{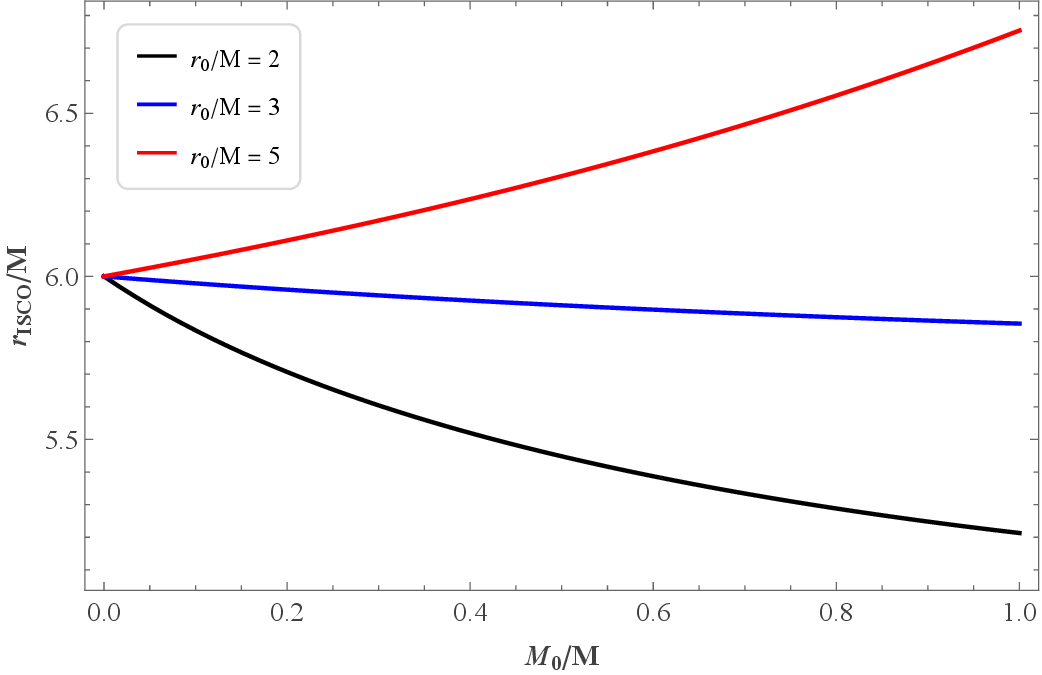}
    \caption{}
    \label{fig:rISCO-r0}
  \end{subfigure}
  \hfill
  \begin{subfigure}[b]{0.32\textwidth}
    \centering
    \includegraphics[width=\linewidth, height=\linewidth]{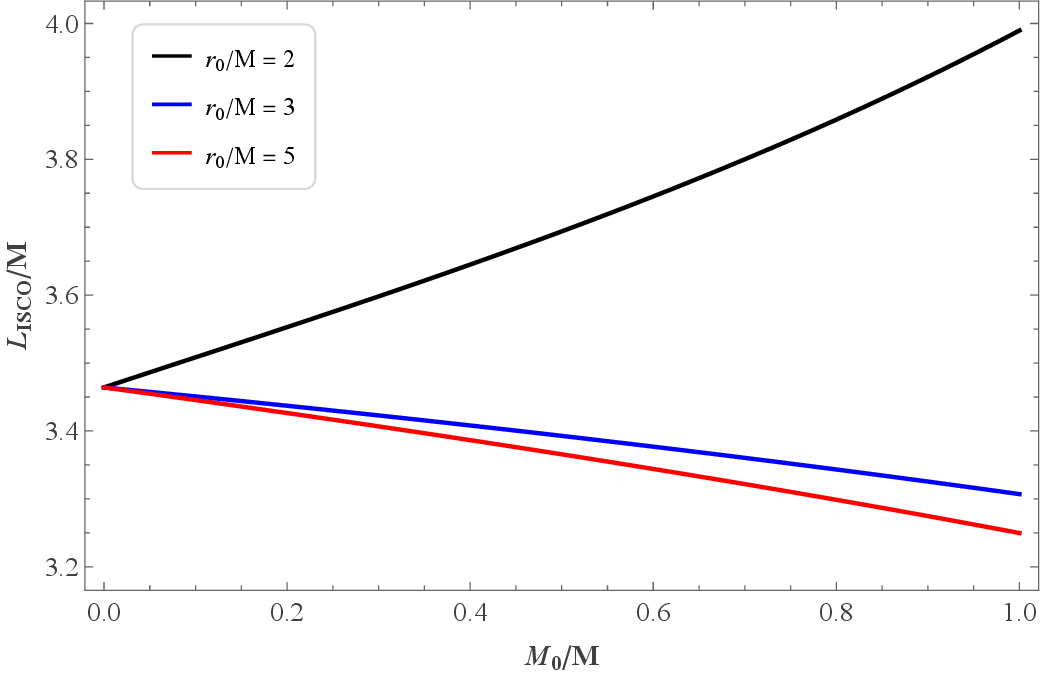}
    \caption{}
    \label{fig:LISCO-r0}
  \end{subfigure}
  \hfill
  \begin{subfigure}[b]{0.32\textwidth}
    \centering
    \includegraphics[width=\linewidth, height=\linewidth]{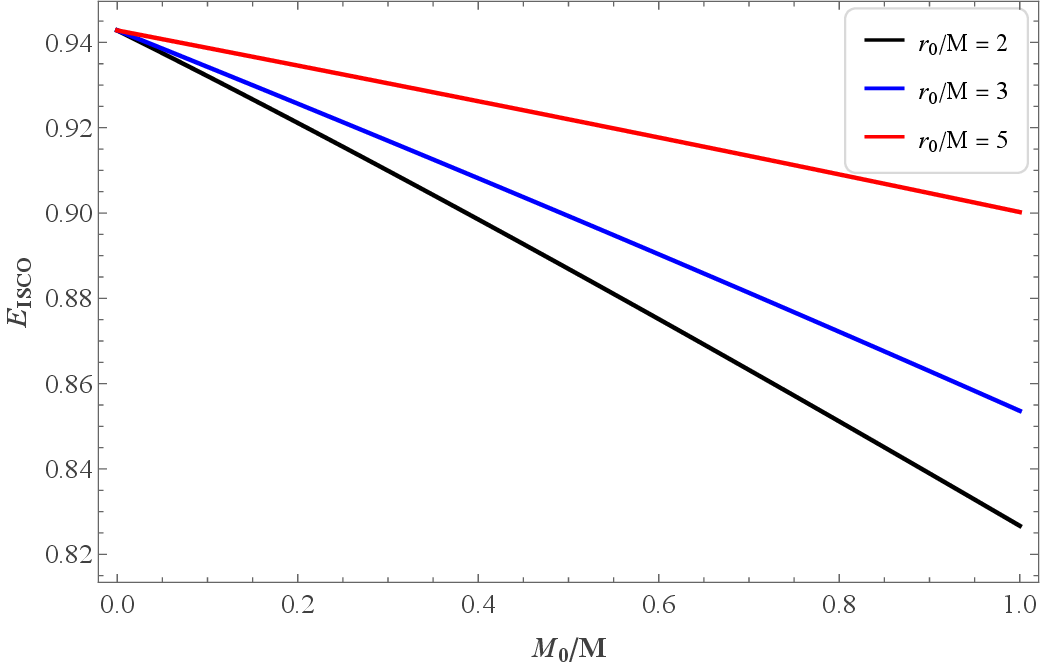}
    \caption{}
    \label{fig:EISCO-r0}
  \end{subfigure}
  
  \caption{ISCO radius, angular momentum, and energy for the Schwarzschild-like black hole embedded in an exponential-sphere dark matter halo. Top row: \(r_{\rm ISCO}/M\) (a), \(L_{\rm ISCO}/M\) (b), and \(E_{\rm ISCO}\) (c) versus \(r_0/M\), for \(M_0/M=0\) (Schwarzschild), \(0.05\), and \(0.1\). Bottom row: \(r_{\rm ISCO}/M\) (d), \(L_{\rm ISCO}/M\) (e), and \(E_{\rm ISCO}\) (f) versus \(M_0/M\), for \(r_0/M=2\), \(3\), and \(5\).}
  \label{fig:ISCO-params}
\end{figure}

Figure~\ref{fig:ISCO-params} shows the same qualitative structure found for the MBO. As a function of $r_0/M$ [panels (a)--(c)], $r_{\rm ISCO}$ and $E_{\rm ISCO}$ are non-monotonic, dipping below their Schwarzschild values around $r_0/M\sim1.5$--$2$ before recovering, while $L_{\rm ISCO}$ decreases monotonically with no such dip. As a function of $M_0/M$ [panels (d)--(f)], the response depends on how extended the halo is: for the more compact halos ($r_0/M=2,3$), $r_{\rm ISCO}$ decreases and $L_{\rm ISCO}$ increases with $M_0$, whereas for the extended halo ($r_0/M=5$) both trends reverse. $E_{\rm ISCO}$ is the only quantity that decreases monotonically with $M_0$ regardless of $r_0$, though more weakly so as the halo becomes more extended. This confirms that the halo mass and its spatial extent act through genuinely different channels, rather than a single effective parameter.

\begin{figure}[htbp]
\centering
\begin{subfigure}[b]{0.48\linewidth}
\centering
\includegraphics[width=\linewidth]{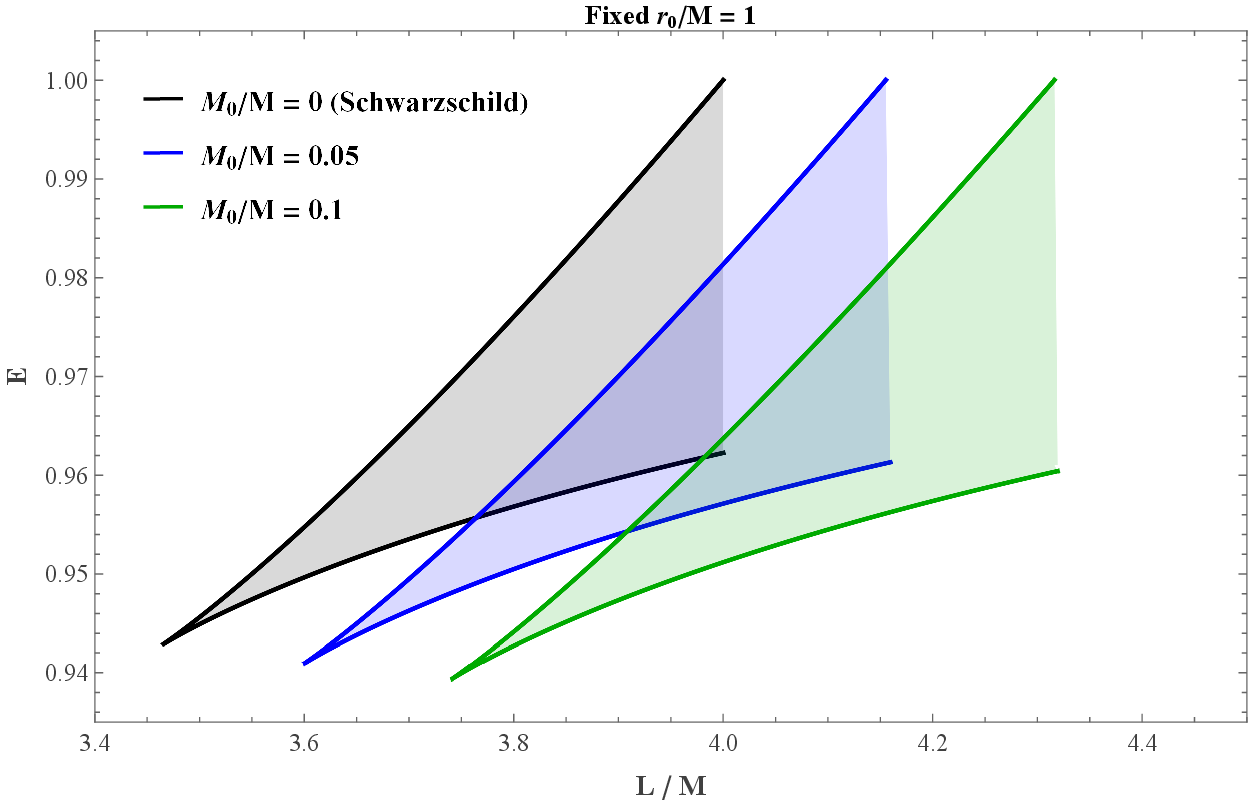}

\label{fig:el_phase_space_M0}
\end{subfigure}
\hfill
\begin{subfigure}[b]{0.48\linewidth}
\centering
\includegraphics[width=\linewidth]{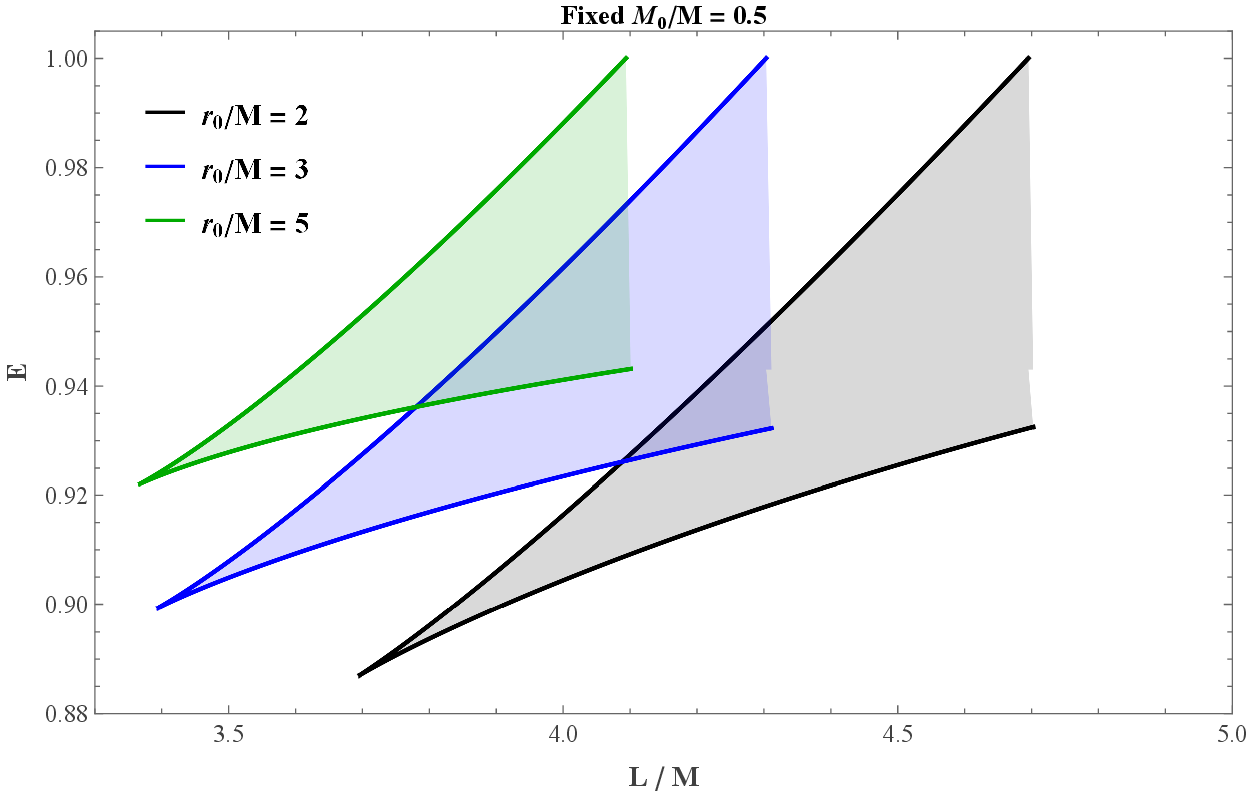}

\label{fig:el_phase_space_r0}
\end{subfigure}
\caption{The energy--angular momentum phase space ($E$--$L$ plane) for massive particles around the Schwarzschild-like black hole embedded in an exponential-sphere dark matter halo. The shaded regions represent bound orbits ($E\le1$), bounded above by the marginally bound orbit ($E=1$) and below by the ISCO. (a) Variation with the halo mass $M_0/M$ at fixed $r_0/M=1$. (b) Variation with the scale radius $r_0/M$ at fixed $M_0/M=0.5$.}
\label{fig:el_phase_space}
\end{figure}
Figure~\ref{fig:el_phase_space} shows the allowed $(E,L)$ region for bound orbits, a lens-shaped strip bounded below by the ISCO and closing at $E=1$ at the MBO. At fixed $r_0/M=1$ [panel (a)], increasing $M_0$ translates the whole region toward larger $L$, with little change in its width. At fixed $M_0/M=0.5$ [panel (b)], increasing $r_0$ instead shifts the region toward smaller $L$, visibly narrows it, and raises its lower boundary to higher $E$. The two halo parameters thus reshape the bound-orbit phase space in different ways, consistent with the distinct trends already seen in the MBO and ISCO themselves.

\section{Periodic Orbits around the Model}
\label{sec:periodic}
Periodic orbits constitute a powerful tool for probing the strong-field dynamics of black holes and for modelling gravitational-wave emission from extreme mass-ratio inspirals (EMRIs). In a static, spherically symmetric spacetime, a test particle on a bound geodesic is characterised by two fundamental frequencies: the radial frequency $\omega_{r}$ and the azimuthal frequency $\omega_{\phi}$. A trajectory is classified as periodic when the ratio of these frequencies takes a rational value, allowing the orbit to close upon itself after a finite number of oscillations. Following the taxonomy introduced by Levin and Perez-Giz, each periodic orbit is uniquely labelled by a triplet of positive integers $(z, w, v)$, where $z$ is the zoom number (denoting the number of radial oscillations required for the orbit to close), $w$ is the whirl number (representing the number of full revolutions around the black hole per radial cycle), and $v$ is the vertex number (specifying the ordering of successive leaves). The frequency ratio is parameterised by the rational number $q$ as
\begin{equation}
q \equiv \frac{\omega_{\phi}}{\omega_{r}} - 1 = w + \frac{v}{z}.
\label{eq:q_definition}
\end{equation}
For given values of the halo parameters $M_{0}/M$ and $r_{0}/M$, periodic orbits are strictly confined to the bound region defined in Section 3, satisfying $E_{\mathrm{ISCO}} \leq E \leq 1$ and $L \geq L_{\mathrm{ISCO}}$. Using the radial and azimuthal equations of motion, $q$ can be expressed explicitly as
\begin{equation}
q = \frac{1}{\pi} \int_{r_{1}}^{r_{2}} \frac{L}{r^{2} \sqrt{E^{2} - V_{\mathrm{eff}}(r)}} \, dr - 1,
\label{eq:q_integral}
\end{equation}
where $r_{1}$ and $r_{2}$ are the turning points (periastron and apastron) determined by the condition $E^{2} = V_{\mathrm{eff}}(r)$. For specified pairs of specific energy $E$ and angular momentum $L$, this integral is evaluated numerically using standard quadrature methods.

To explore the bound parameter space systematically, we fix one orbital constant to a representative average value, $L_{\rm av} = (L_{\rm MBO}+L_{\rm ISCO})/2$ or $E_{\rm av} = (1+E_{\rm ISCO})/2$ (recalling $E_{\rm MBO}=1$), and solve for the other. Solving $q(E)=w+v/z$ at $L=L_{\rm av}$ gives the periodic-orbit energies listed in Table~\ref{tab:periodic_energies_esm} (versus $M_0/M$ at fixed $r_0/M=1$) and Table~\ref{tab:periodic_energies_esm_r0} (versus $r_0/M$ at fixed $M_0/M=0.5$). Solving $q(L)=w+v/z$ at $E=E_{\rm av}$ gives the corresponding angular momenta in Table~\ref{tab:periodic_L_esm_M0} and Table~\ref{tab:periodic_L_esm_r0}, for the same two parameter scans.

\begin{table}[htbp]
    \centering
    \setlength{\arrayrulewidth}{0.5pt}
    \renewcommand{\arraystretch}{1.2}
    \scriptsize
    \setlength{\tabcolsep}{3pt}
    \resizebox{\textwidth}{!}{%
    \begin{tabular}{cccccccccc}
    \toprule
    \(M_0/M\) & \(L_{\text{av}}\) & \(E(1,1,0)\) & \(E(1,2,0)\) & \(E(2,1,1)\) & \(E(2,2,1)\) & \(E(3,1,2)\) & \(E(3,2,2)\) & \(E(4,1,3)\) & \(E(4,2,3)\) \\
    \midrule
    0.000000 & 3.732051 & 0.965425 & 0.968383 & 0.968027 & 0.968434 & 0.968225 & 0.968438 & 0.968285 & 0.968440 \\
    0.050000 & 3.877008 & 0.963903 & 0.967422 & 0.966990 & 0.967486 & 0.967229 & 0.967491 & 0.967302 & 0.967493 \\
    0.100000 & 4.028212 & 0.962490 & 0.966648 & 0.966128 & 0.966727 & 0.966414 & 0.966734 & 0.966502 & 0.966736 \\
    \bottomrule
    \end{tabular}%
    }%
    \caption{Periodic-orbit energies $E(z,w,v)$ as functions of the halo mass $M_0/M$, at fixed $r_0/M=1$.}
\label{tab:periodic_energies_esm}
\end{table}

\begin{table}[htbp]
    \centering
    \setlength{\arrayrulewidth}{0.5pt}
    \renewcommand{\arraystretch}{1.2}
    \scriptsize
    \setlength{\tabcolsep}{3pt}
    \resizebox{\textwidth}{!}{%
    \begin{tabular}{cccccccccc}
    \toprule
    \(r_0/M\) & \(L_{\text{av}}\) & \(E(1,1,0)\) & \(E(1,2,0)\) & \(E(2,1,1)\) & \(E(2,2,1)\) & \(E(3,1,2)\) & \(E(3,2,2)\) & \(E(4,1,3)\) & \(E(4,2,3)\) \\
    \midrule
    2.000000 & 4.194561 & 0.936200 & 0.938159 & 0.937951 &0.938184 & 0.938071 & 0.938186 & 0.938105 & 0.938186 \\
    3.000000 & 3.848127 & 0.942672 & 0.943756 & 0.943643 & 0.943769 & 0.943708 & 0.943770 & 0.943727 & 0.943770 \\
    5.000000 & 3.729797 & 0.954787 & 0.956178 & 0.956021 & 0.956198 & 0.956110 & 0.956200 & 0.956136 & 0.956200 \\
    \bottomrule
    \end{tabular}%
    }%
    \caption{Periodic-orbit energies $E(z,w,v)$ as functions of $r_0/M$, at fixed $M_0/M=0.5$.}
\label{tab:periodic_energies_esm_r0}
\end{table}

\begin{table}[htbp]
    \centering
    \setlength{\arrayrulewidth}{0.5pt}
    \renewcommand{\arraystretch}{1.2}
    \scriptsize
    \setlength{\tabcolsep}{3pt}
    \resizebox{\textwidth}{!}{%
    \begin{tabular}{cccccccccc}
    \toprule
    \(M_0/M\) & \(E_{\text{av}}\) & \(L(1,1,0)\) & \(L(1,2,0)\) & \(L(2,1,1)\) & \(L(2,2,1)\) & \(L(3,1,2)\) & \(L(3,2,2)\) & \(L(4,1,3)\) & \(L(4,2,3)\) \\
    \midrule
    0.000000 & 0.971405 & 3.784078 & 3.759377 & 3.762350 & 3.758960 & 3.760687 & 3.758927 & 3.760186 & 3.758917 \\
    0.050000 & 0.970439 & 3.933728 & 3.904433 & 3.908051 & 3.903908 & 3.906040 & 3.903866 & 3.905428 & 3.903853 \\
    0.100000 & 0.969658 & 4.090405 & 4.055809 & 4.060208 & 4.055150 & 4.057780 & 4.055095 & 4.057035 & 4.055078 \\
    \bottomrule
    \end{tabular}%
    }%
  \caption{Periodic-orbit angular momenta $L(z,w,v)$ as functions of the halo mass $M_0/M$, at fixed $r_0/M=1$.}
\label{tab:periodic_L_esm_M0}
\end{table}

\begin{table}[htbp]
    \centering
    \setlength{\arrayrulewidth}{0.5pt}
    \renewcommand{\arraystretch}{1.2}
    \scriptsize
    \setlength{\tabcolsep}{3pt}
    \resizebox{\textwidth}{!}{%
    \begin{tabular}{cccccccccc}
    \toprule
    \(r_0/M\) & \(E_{\text{av}}\) & \(L(1,1,0)\) & \(L(1,2,0)\) & \(L(2,1,1)\) & \(L(2,2,1)\) & \(L(3,1,2)\) & \(L(3,2,2)\) & \(L(4,1,3)\) & \(L(4,2,3)\) \\
    \midrule
    2.000000 & 0.943451 & 4.257131 & 4.239983 & 4.241741 & 4.239778 & 4.240723 & 4.239764 & 4.240430 & 4.239760 \\
    3.000000 & 0.949643 & 3.908390 & 3.899243 & 3.900162 & 3.899139 & 3.899626 & 3.899132 & 3.899474 & 3.899130 \\
    5.000000 & 0.960988 & 3.783918 & 3.772790 & 3.774018 & 3.772637 & 3.773316 & 3.772626 & 3.773111 & 3.772623 \\
    \bottomrule
    \end{tabular}%
    }%
  \caption{Periodic-orbit angular momenta $L(z,w,v)$ as functions of $r_0/M$, at fixed $M_0/M=0.5$.}
\label{tab:periodic_L_esm_r0}
\end{table}

\begin{figure}[htbp]
\centering
\begin{subfigure}[b]{0.48\linewidth}
\centering
\includegraphics[width=\linewidth]{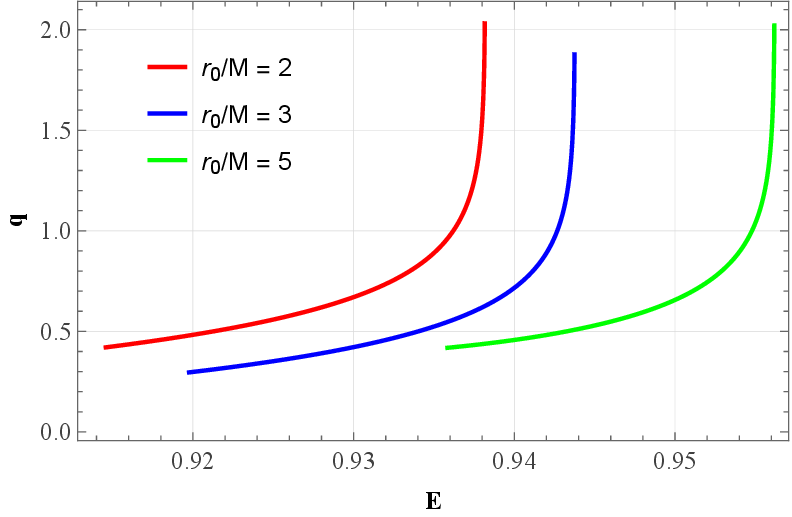}
\label{fig:q-E-r0}
\end{subfigure}
\hfill
\begin{subfigure}[b]{0.48\linewidth}
\centering
\includegraphics[width=\linewidth]{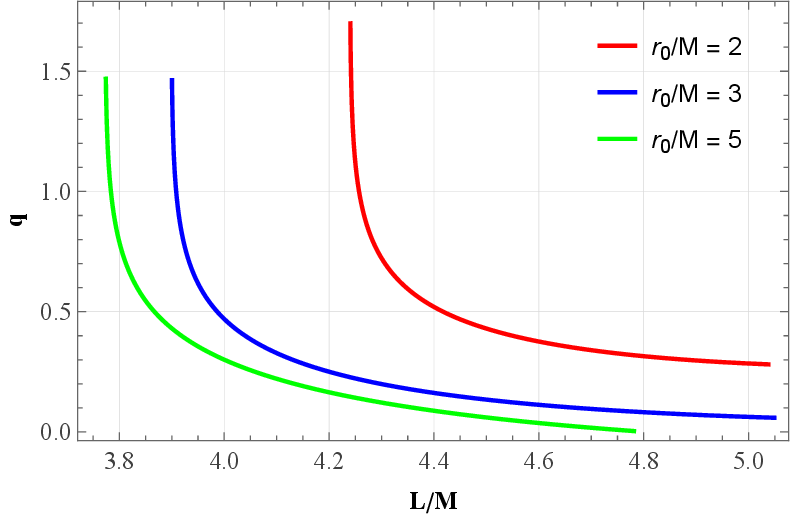}
\label{fig:q-L-r0}
\end{subfigure}

\vspace{0.5cm}

\begin{subfigure}[b]{0.48\linewidth}
\centering
\includegraphics[width=\linewidth]{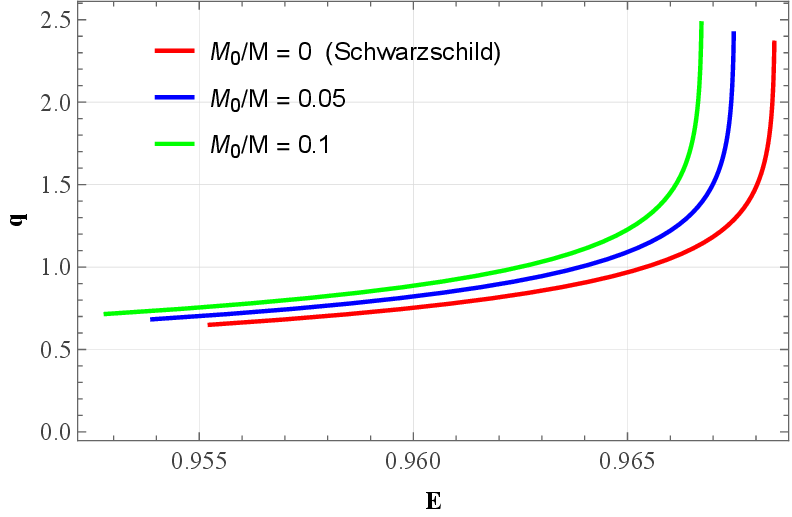}
\label{fig:q-E-M0}
\end{subfigure}
\hfill
\begin{subfigure}[b]{0.48\linewidth}
\centering
\includegraphics[width=\linewidth]{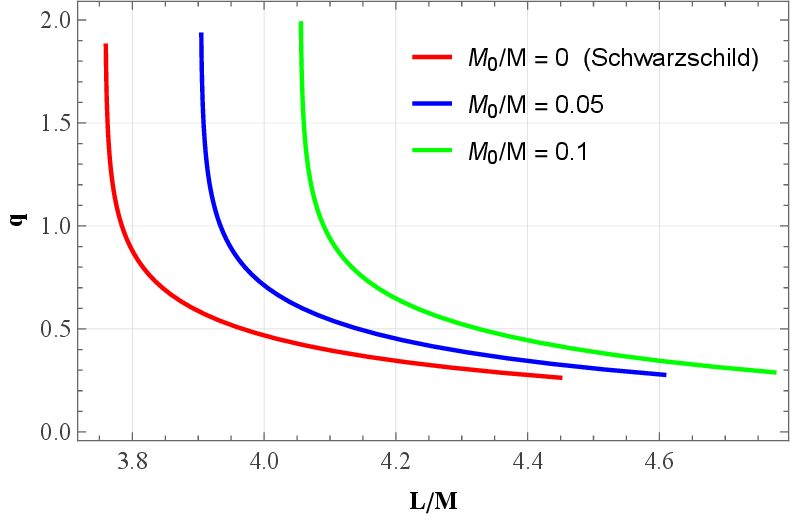}
\label{fig:q-L-M0}
\end{subfigure}

\caption{The dependence of the rational frequency ratio \(q = w+v/z\) on energy \(E\) (left) and angular momentum \(L/M\) (right) for periodic orbits in the ESM black hole. Top row: fixed \(M_0/M=0.5\) and varying \(r_0/M=2,3,5\). Bottom row: fixed \(r_0/M=1\) and varying \(M_0/M=0,0.05,0.1\). In the left panels, \(L=L_{\rm av}\); in the right panels, \(E=E_{\rm av}\). The curves reveal the distinct and opposite effects of the halo mass and its scale radius on the periodic-orbit spectrum.}
\label{fig:q-vs-EL-r0-M0}
\end{figure}

Figure~\ref{fig:q-vs-EL-r0-M0} shows that \(r_0\) and \(M_0\) leave clearly different imprints on the periodic-orbit spectrum. Varying \(r_0\) at fixed \(M_0/M=0.5\) [top row] shifts the \(q(E)\) and \(q(L)\) curves substantially and in opposite directions on the two axes: larger \(r_0\) moves the curve to higher \(E\) but lower \(L\), with little overlap between the three cases. Varying \(M_0\) over the range \(M_0/M=0, 0.05, 0.1\) at fixed \(r_0/M=1\) [bottom row] produces a weaker effect: the \(q(E)\) curves nearly coincide, separating only near their respective divergences, while the \(q(L)\) curves are shifted by a smaller but still resolved amount. Here, larger \(M_0\) moves both curves to higher values of \(E\) and \(L\). This distinct asymmetry reinforces the conclusion that the halo mass and its spatial extent govern the geodesic dynamics through fundamentally different physical channels.

\begin{figure}[htbp]
\centering
\includegraphics[width=0.6\linewidth]{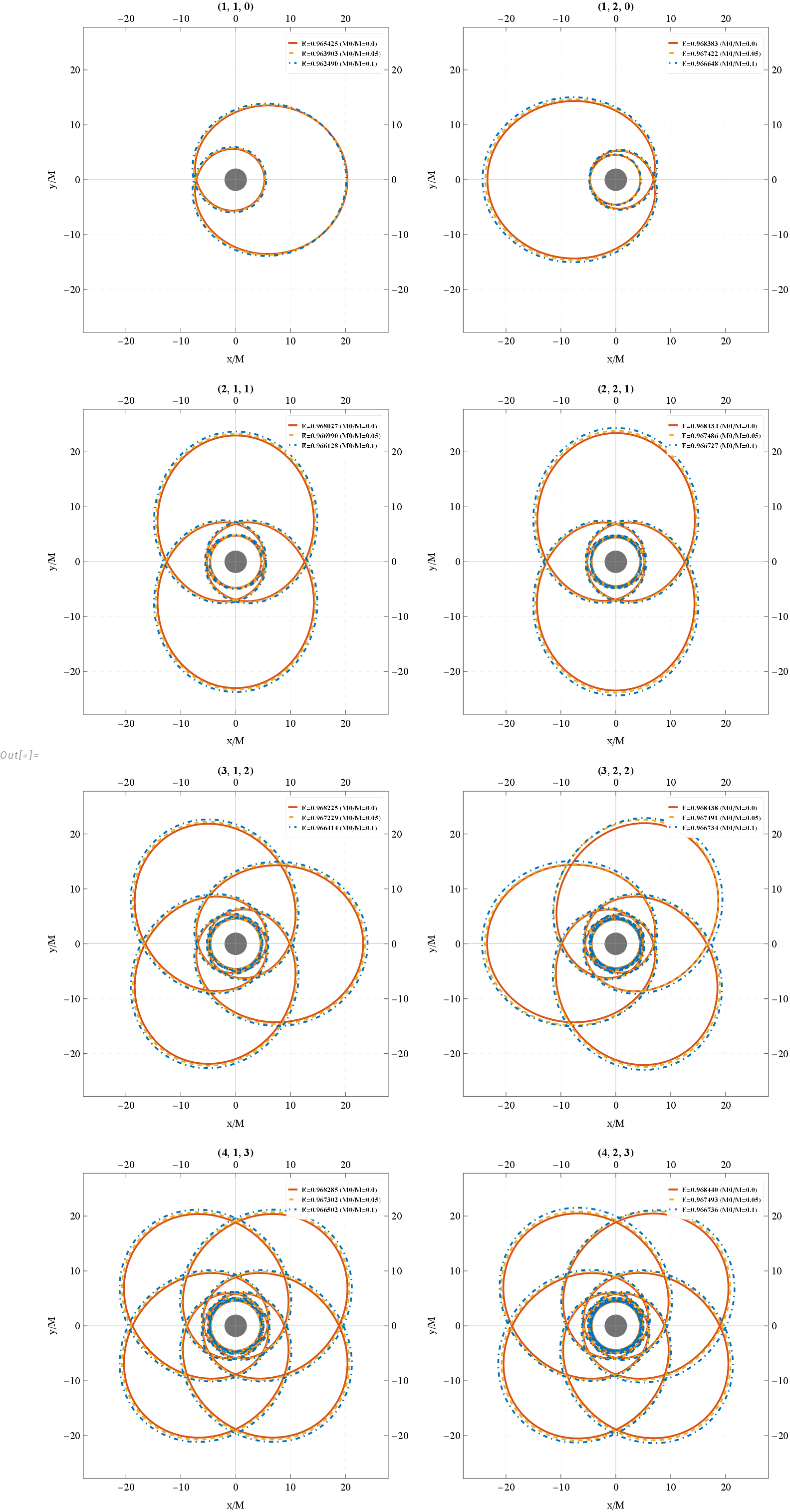}
\caption{Periodic orbits with rational frequency ratio $q = w + v/z$ around a Schwarzschild-like black hole embedded in an exponential-sphere dark matter halo, for $r_0/M = 1$ and fixed angular momentum $L = L_{\rm av}$. Solid, dashed, and dot-dashed curves correspond to $M_0/M = 0.0$, $0.05$, and $0.1$, respectively; the gray disk marks the event horizon. Increasing $M_0$ visibly reshapes the orbital geometry relative to the Schwarzschild case ($M_0/M=0.0$).}
\label{fig:periodic_orbits_ESM}
\end{figure}

\begin{figure}[htbp]
\centering
\includegraphics[width=0.6\linewidth]{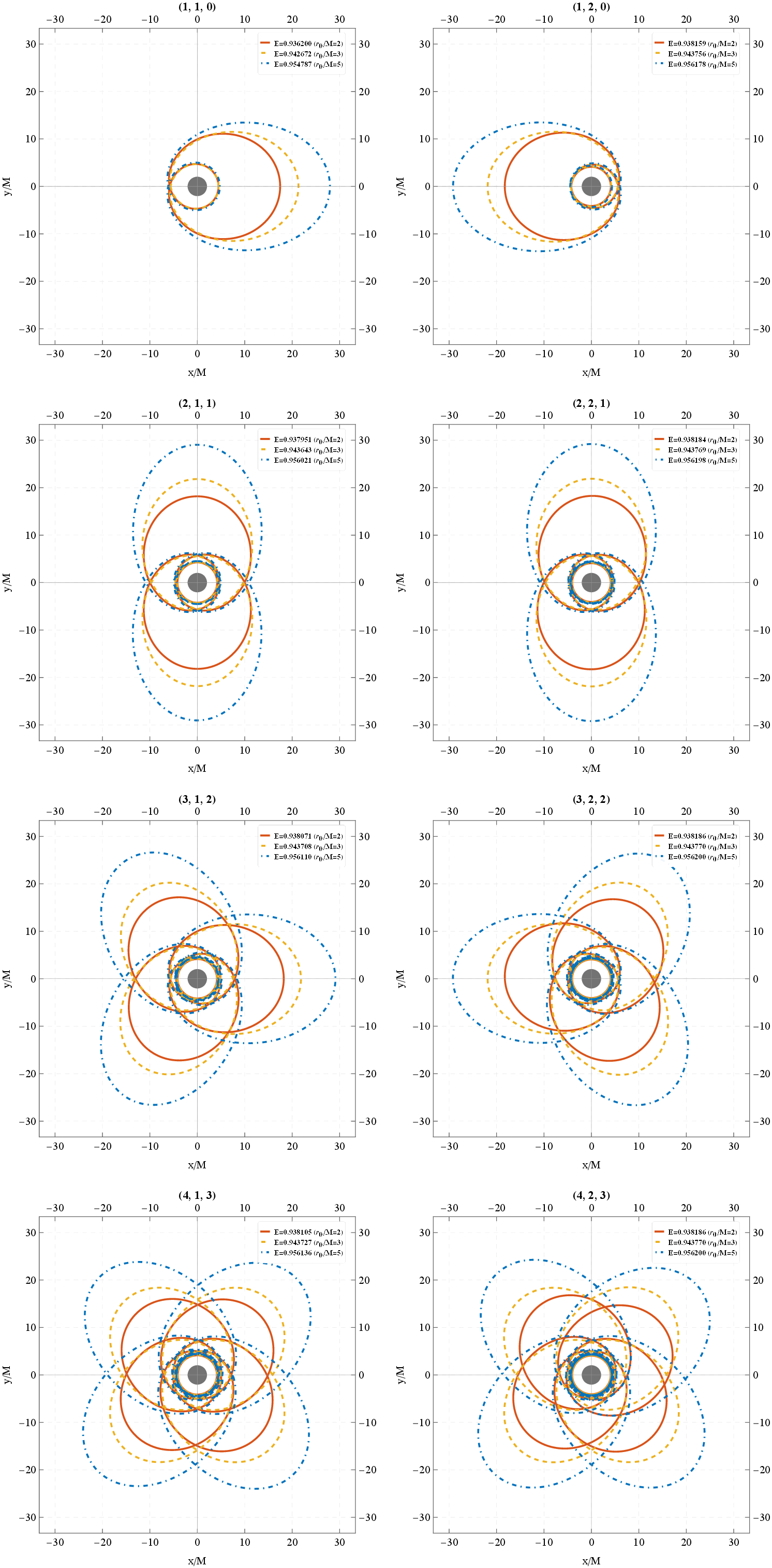}
\caption{Periodic orbits with rational frequency ratio $q = w + v/z$ around a Schwarzschild-like black hole embedded in an exponential-sphere dark matter halo, for fixed halo mass $M_0/M = 0.5$ and fixed angular momentum $L = L_{\rm av}$. Solid, dashed, and dot-dashed curves correspond to $r_0/M = 2$, $3$, and $5$, respectively; the gray disk marks the event horizon. Increasing $r_0$ visibly reshapes the orbital geometry.}
\label{fig:periodic_orbits_ESM_r0}
\end{figure}

\begin{figure}[htbp]
\centering
\includegraphics[width=0.6\linewidth]{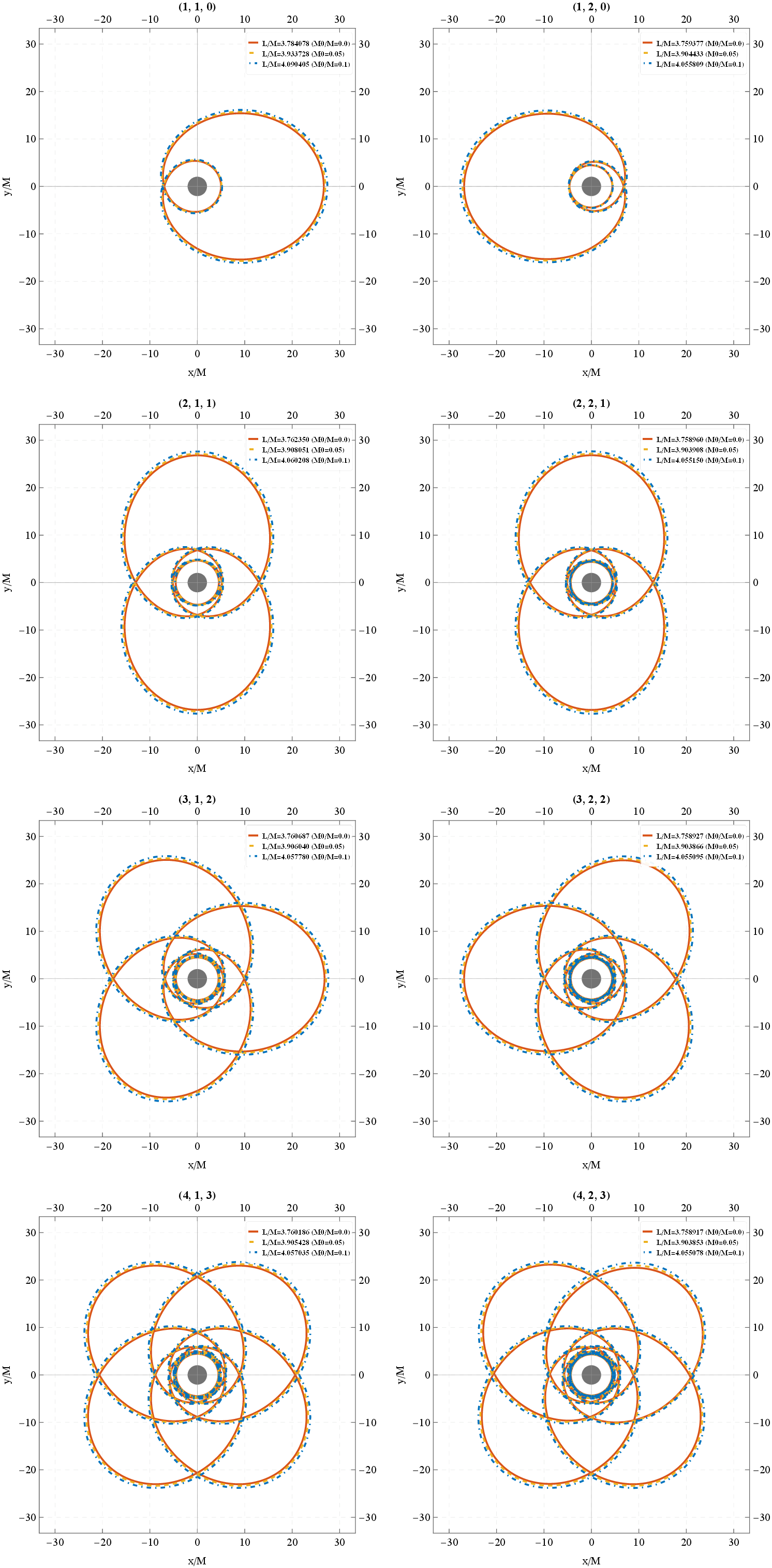}
\caption{Periodic orbits with rational frequency ratio $q = w + v/z$ around a Schwarzschild-like black hole embedded in an exponential-sphere dark matter halo, for fixed halo scale radius $r_0/M = 1$ and fixed energy $E = E_{\rm av}$. Solid, dashed, and dot-dashed curves correspond to $M_0/M = 0.0$, $0.05$, and $0.1$, respectively; the gray disk marks the event horizon. Increasing $M_0$ visibly reshapes the orbital geometry.}
\label{fig:periodic_orbits_ESM_E}
\end{figure}

\begin{figure}[htbp]
\centering
\includegraphics[width=0.6\linewidth]{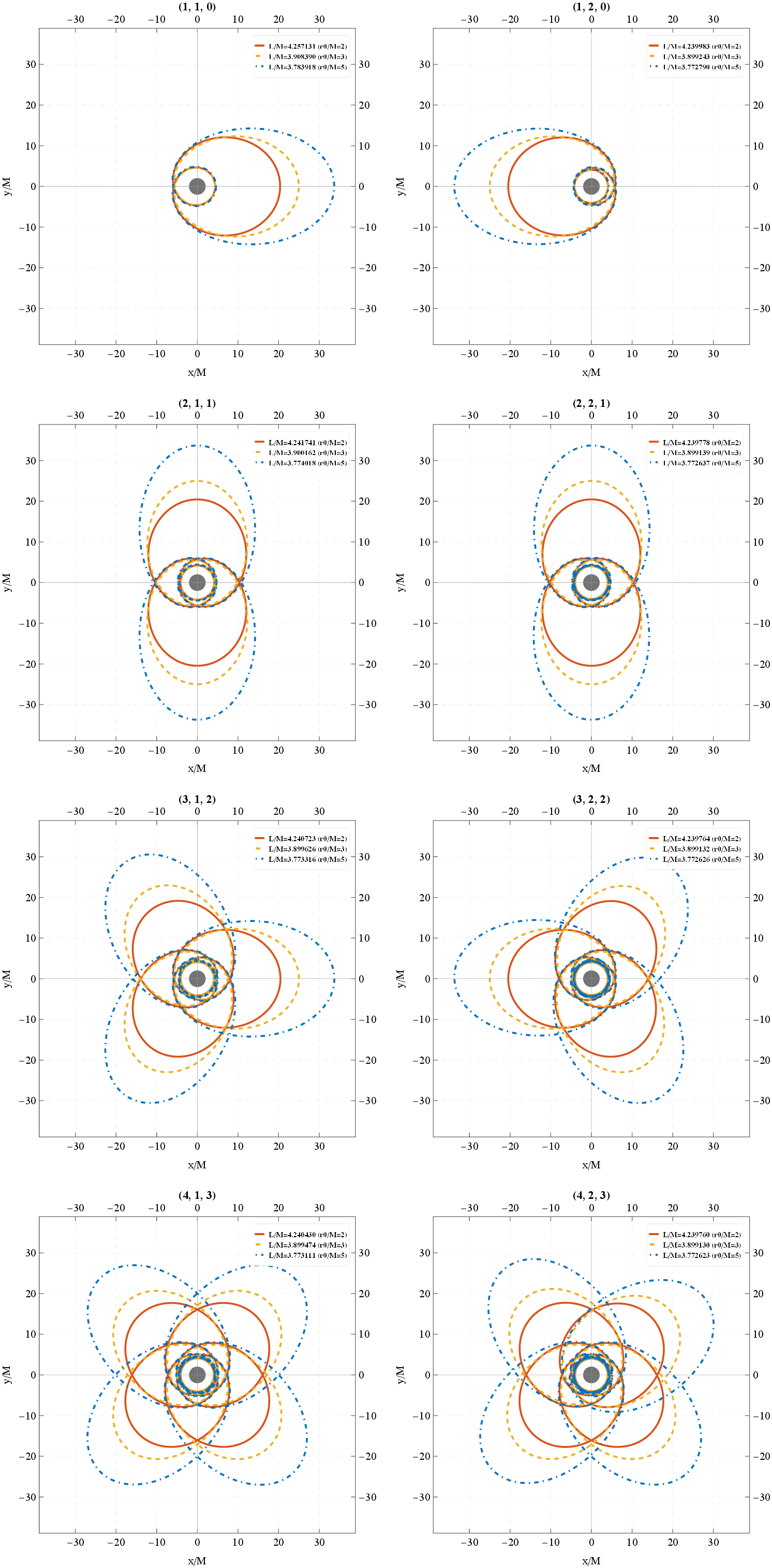}
\caption{Periodic orbits with rational frequency ratio $q = w + v/z$ around a Schwarzschild-like black hole embedded in an exponential-sphere dark matter halo, for fixed halo mass $M_0/M = 0.5$ and fixed energy $E = E_{\rm av}$. Solid, dashed, and dot-dashed curves correspond to $r_0/M = 2$, $3$, and $5$, respectively; the gray disk marks the event horizon. Increasing $r_0$ visibly reshapes the orbital geometry.}
\label{fig:periodic_orbits_ESM_r0E}
\end{figure}

Figures~\ref{fig:periodic_orbits_ESM}--\ref{fig:periodic_orbits_ESM_r0E} display the periodic orbits, classified by the rational ratio $q = w + v/z$, around the ESM-dressed black hole. To isolate the effects of the halo mass $M_0$ and scale radius $r_0$, we examine the trajectories under two distinct constraints: fixed average angular momentum $L_{\rm av}$ [Figs.~\ref{fig:periodic_orbits_ESM} and \ref{fig:periodic_orbits_ESM_r0}] and fixed average energy $E_{\rm av}$ [Figs.~\ref{fig:periodic_orbits_ESM_E} and \ref{fig:periodic_orbits_ESM_r0E}].

At fixed $L_{\rm av}$, the halo parameters induce distinct geometric deformations. Increasing $M_0$ at $r_0/M = 1$ (Fig.~\ref{fig:periodic_orbits_ESM}) deepens the potential well, causing a slight inward compaction of the orbits to preserve the fixed angular momentum; this effect remains subtle for small $M_0$ but reflects a tighter binding. Conversely, increasing $r_0$ at $M_0/M = 0.5$ (Fig.~\ref{fig:periodic_orbits_ESM_r0}) distributes the halo mass over a larger volume, flattening the inner potential. This spatial redistribution expands the trajectories, allowing the same $L_{\rm av}$ to support a larger apoapsis.

At fixed $E_{\rm av}$, the orbital response is governed by the corresponding shift in the required $L$. Increasing $M_0$ at $r_0/M = 1$ (Fig.~\ref{fig:periodic_orbits_ESM_E}) necessitates a larger $L$ to sustain the same rational $q$, leading to a pronounced outward expansion of the zoom-whirl topology. In contrast, increasing $r_0$ at $M_0/M = 0.5$ (Fig.~\ref{fig:periodic_orbits_ESM_r0E}) lowers the required $L$, forcing the particle into a more compact orbit to maintain the fixed energy. These opposing responses confirm that $M_0$ and $r_0$ act through non-degenerate dynamical channels, establishing the periodic-orbit spectrum as a sensitive probe of both the mass and spatial distribution of the dark matter halo.

Having established how the dark matter halo reshapes the periodic orbits, we now examine its imprint on the emitted radiation. Because these trajectories drive the inspiral phase of EMRIs, the orbital deformations induced by $M_0$ and $r_0$ directly modulate the emitted signal. In the following section, we employ the numerical kludge framework to construct the time-domain polarizations $h_+$ and $h_\times$, translating these geometric shifts into observable gravitational-wave signatures.

\section{Gravitational Wave}\label{sec:waveforms}
The gravitational radiation emitted by a stellar-mass compact object moving along the periodic trajectories identified in Section 4 is examined in this section. The system is modelled as an extreme mass-ratio inspiral (EMRI), composed of a compact object of mass $m$ orbiting a supermassive ESM primary black hole of mass $M$. The waveform calculation is performed within the numerical kludge framework, which combines the numerically integrated geodesic trajectory with the standard quadrupole formula. In the adiabatic limit, the radiation-reaction timescale is assumed to be much longer than the orbital period; consequently, the orbital energy and angular momentum are treated as conserved over a single radial cycle, and the backreaction on the orbital motion is neglected.
Within this framework, the metric perturbation in the wave zone is derived from the symmetric trace-free (STF) mass quadrupole tensor, given by
\begin{equation}
h_{ij} = \frac{4 \mu M}{D_{L}} \left( v_{i} v_{j} - \frac{m}{M} n_{i} n_{j} \right),
\label{eq:metric_perturbation}
\end{equation}
where $\mu = Mm/(M + m)^{2}$ is the symmetric mass ratio, $D_{L}$ is the luminosity distance to the source, and $n_{i}$ and $v_{i}$ represent the radial unit vector and the velocity components of the orbiting body, respectively. To obtain the observable waveform, this metric perturbation is projected onto a transverse-traceless frame adapted to the detector. Introducing the orthonormal basis vectors
\begin{align}
e_{X} &= (\cos \zeta, -\sin \zeta, 0), \label{eq:eX}\\
e_{Y} &= (\cos \iota \sin \zeta, \cos \iota \cos \zeta, -\sin \iota), \label{eq:eY}\\
e_{Z} &= (\sin \iota \sin \zeta, \sin \iota \cos \zeta, \cos \iota), \label{eq:eZ}
\end{align}
where $\iota$ denotes the orbital inclination and $\zeta$ specifies the orientation of the orbit relative to the observer, the two independent polarization modes are expressed as
\begin{align}
h_{+} &= -\frac{2 \mu M^{2}}{D_{L} r} \left( 1 + \cos^{2} \iota \right) \cos(2\phi + 2\zeta), \label{eq:hplus}\\
h_{\times} &= -\frac{4 \mu M^{2}}{D_{L} r} \cos \iota \sin(2\phi + 2\zeta), \label{eq:hcross}
\end{align}
with $r$ and $\phi$ denoting the instantaneous orbital radius and phase. These quantities are obtained directly from the numerical integration of the timelike geodesic equations governed by the ESM metric. For the numerical simulations, geometrized units are adopted, with the mass scales fixed at $M \sim 10^{7} M_{\odot}$ and $m \sim 10 M_{\odot}$, and the source placed at a luminosity distance of $D_{L} = 200$ Mpc. The orientation angles are fixed at $\iota = \zeta = \pi/4$.

\begin{figure}[htbp]
\centering
\includegraphics[width=0.85\linewidth]{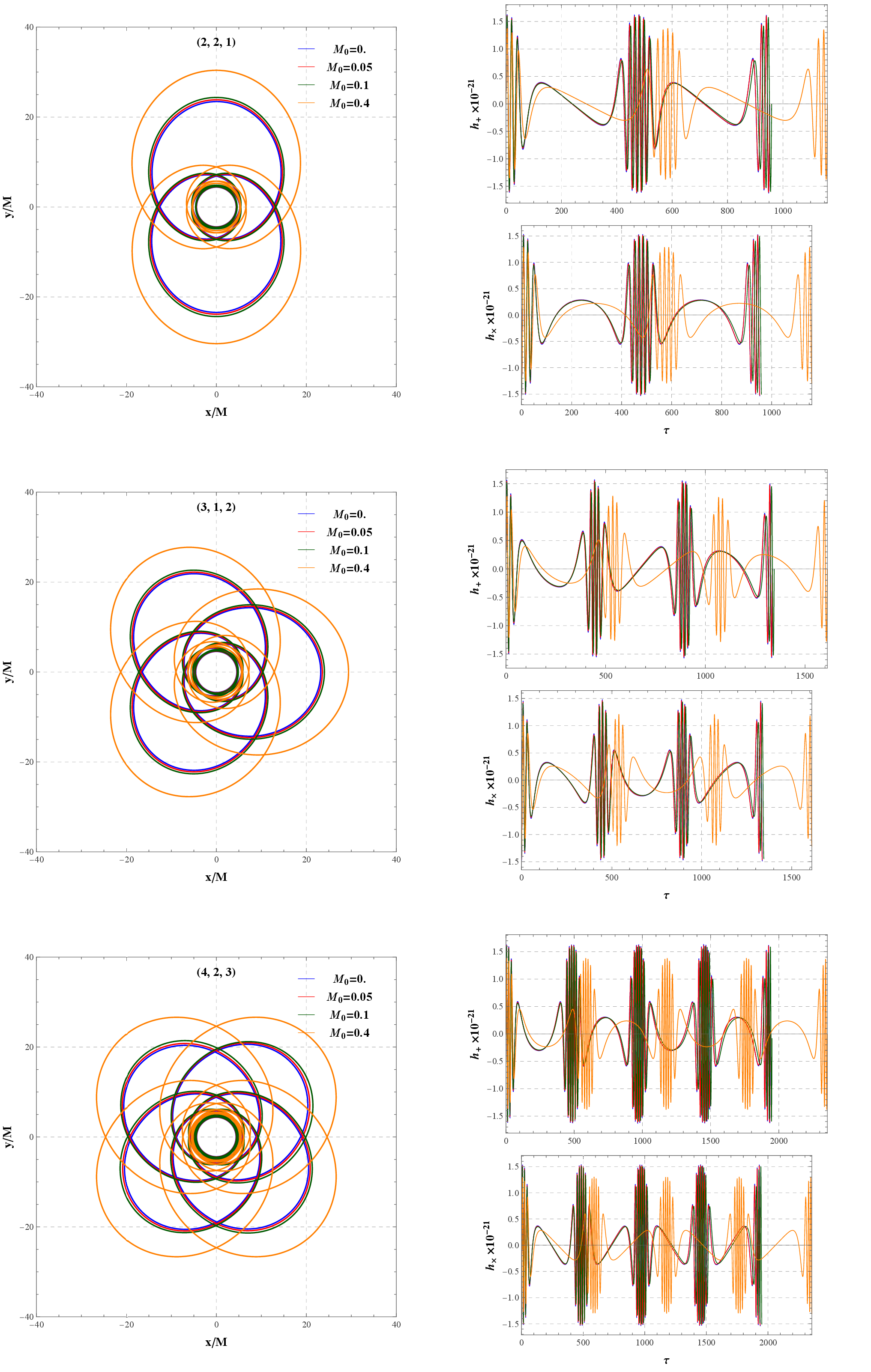}
\caption{Periodic orbits and gravitational-wave polarizations for the $(2,2,1)$,
$(3,1,2)$, and $(4,2,3)$ configurations (top to bottom) around the ESM black
hole, at fixed $r_{0}=1$ and $M_{0}=0,\,0.05,\,0.1,\,0.4$ (blue, red, green,
orange). $M_{0}=0.4$ is added to make the halo-mass effect visible: the other
three curves remain nearly Schwarzschild-like, while $M_{0}=0.4$ clearly
separates in orbital extent and whirl-burst dephasing.}
\label{fig:GW_M0scan}
\end{figure}

\begin{figure}[htbp]
\centering
\includegraphics[width=0.85\linewidth]{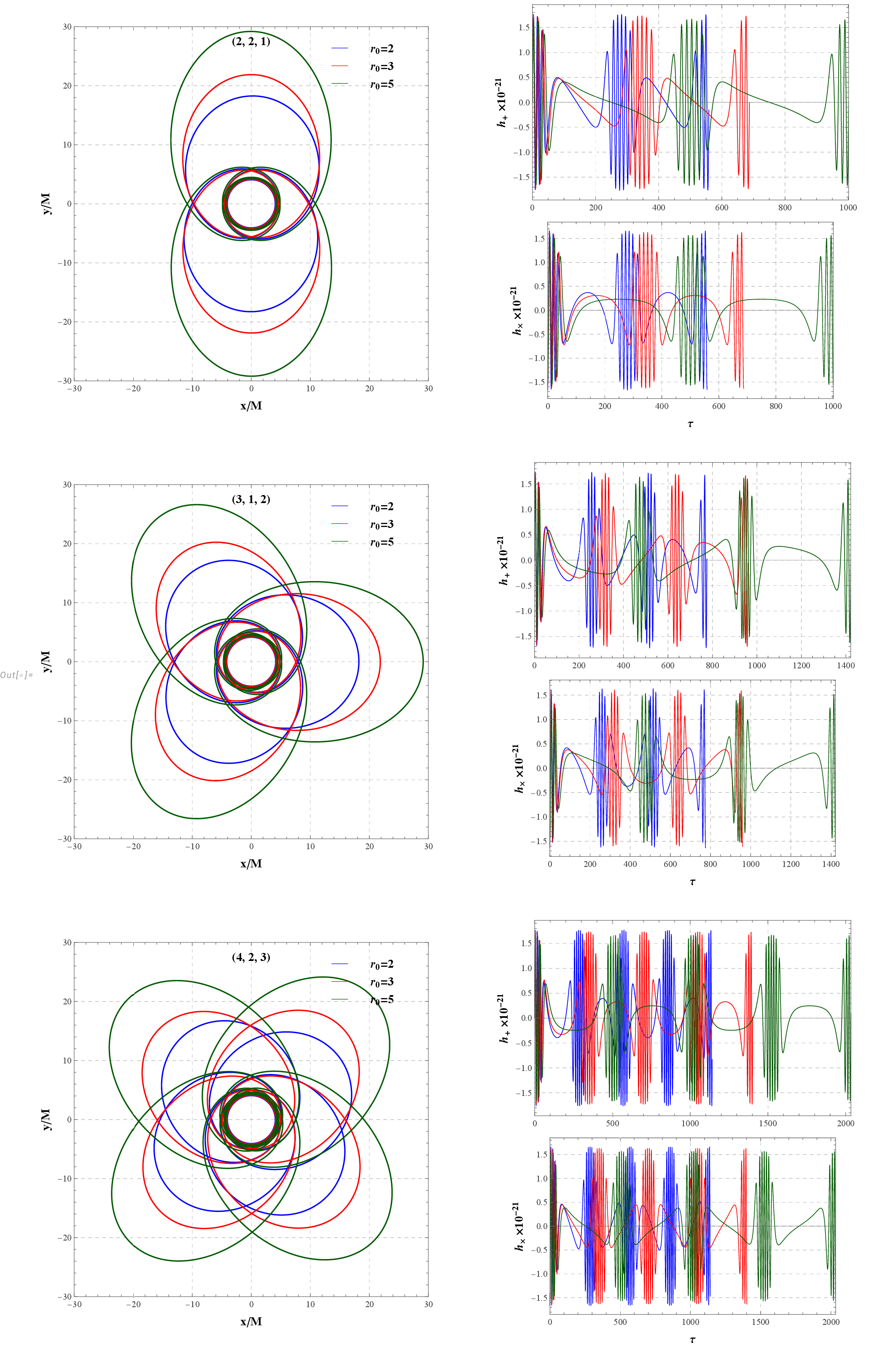}
\caption{Periodic orbits and gravitational-wave polarizations for the $(2,2,1)$,
$(3,1,2)$, and $(4,2,3)$ configurations (top to bottom) around the ESM black
hole, at fixed $M_{0}=0.5$ and $r_{0}=2,\,3,\,5$ (blue, red, green).}
\label{fig:GW_r0scan}
\end{figure}

Figures~\ref{fig:GW_M0scan} and \ref{fig:GW_r0scan} illustrate the periodic
orbital trajectories in the equatorial plane ($x/M$--$y/M$) and the
corresponding gravitational-wave (GW) polarizations ($h_+$ and $h_\times$)
emitted by a test particle in an extreme mass-ratio inspiral (EMRI) around a
Schwarzschild-like black hole embedded in an exponential dark matter halo. The
analysis is separated into two distinct parametric regimes: varying the halo
mass $M_0$ at fixed scale radius ($r_0/M=1$), and varying the scale radius $r_0$
at fixed halo mass ($M_0/M=0.5$). Representative periodic orbits characterized
by the rational number $q=w+v/z$, corresponding to the configurations
$(2,2,1)$, $(3,1,2)$, and $(4,2,3)$, are shown in both cases.

\subsection{Impact of Halo Mass ($M_0$)}

In Fig.~\ref{fig:GW_M0scan}, we fix $r_0/M=1$ and examine the effect of the
halo mass $M_0\in\{0.0,0.05,0.1,0.4\}$, where $M_0=0.0$ recovers the standard
Schwarzschild vacuum geometry. For the two smallest increments, $M_0=0.05$ and
$0.1$, the orbital trajectories and waveforms remain nearly indistinguishable
from the Schwarzschild case, with the corresponding curves overlapping almost
entirely in both panels. Once $M_0$ is increased to $0.4$, however, a
pronounced departure emerges: the periodic orbit expands considerably, with the
outer zoom excursions reaching a maximal radius of order $30\,M$, compared to
roughly $23$--$24\,M$ for $M_0\leq0.1$. This enlargement reflects the growth of
the total mass enclosed within the orbit as $M_0$ increases, which acts, to
leading order, in analogy with increasing the mass of an isolated Schwarzschild
black hole. The associated average angular momentum $L_{\rm av}$ grows
accordingly, from $3.73$ at $M_0=0$ to $5.07$ at $M_0=0.4$, consistent with the
visibly larger orbital span seen in the left panels.

This orbital enlargement is directly imprinted on the emitted waveforms. All
four curves share a nearly identical initial burst near $\tau\simeq0$, since the
first periapsis passage is only weakly sensitive to $M_0$; beyond this point,
however, the $M_0=0.4$ signal quickly departs from the other three. Its zoom
phases last visibly longer, its whirl bursts are pushed to progressively later
values of $\tau$, and by the second or third burst the $M_0=0.4$ waveform is
completely out of phase with the near-degenerate $M_0=0,0.05,0.1$ triplet. This
effect is most striking in the $(4,2,3)$ configuration, where the longer
integration window makes the accumulated dephasing unambiguous. Peak amplitudes
remain comparable across all four cases, indicating that, in this regime, it is
primarily the timing and phase structure of the zoom-whirl cycle -- rather than
the instantaneous strength of the signal -- that carries the imprint of $M_0$.

\subsection{Impact of Scale Radius ($r_0$)}

Figure~\ref{fig:GW_r0scan} isolates the effect of the halo's spatial extent
by fixing $M_0/M=0.5$ and varying $r_0\in\{2,3,5\}$. The scale radius $r_0$
controls the radial distribution of the dark matter density: a larger $r_0$
corresponds to a more diffuse, extended halo, while a smaller $r_0$
concentrates the same total mass closer to the central black hole.

The orbital trajectories show a clear, monotonic dependence on $r_0$: the
periodic orbit expands as $r_0$ increases, with the $r_0=5$ trajectory visibly
more extended than the $r_0=2$ case for all three orbital configurations
considered. This spatial trend is mirrored in the time-domain waveforms, whose
radial period lengthens with increasing $r_0$; the whirl bursts of the $r_0=5$
signal occur at systematically later $\tau$ than those of the $r_0=2$ case,
producing a progressive dephasing similar in appearance to, though driven by a
distinct combination of halo parameters than, the dephasing observed for large
$M_0$. We note that this trend is not a simple monotonic interpolation between
the Schwarzschild limit and a point-mass-halo limit, and a more detailed account
of how $r_0$ reshapes the effective potential at fixed $M_0$ is left for future
work.

\subsection{Physical Interpretation}

Taken together, these results show that the ambient dark matter environment
leaves a clear, and in principle separable, imprint on the emitted
gravitational radiation. Both $M_0$ and $r_0$ act through the same effective
mass function $A(r)$, yet their signatures over the parameter ranges considered
here are not identical: small variations in $M_0$ ($\lesssim0.1$) leave the
waveform essentially unchanged, whereas comparable variations in $r_0$ already
produce a clearly resolved shift in orbital period and burst timing. Only once
$M_0$ becomes a non-negligible fraction of the central black hole mass does its
effect on the periodic-orbit structure and on the emitted waveform become
comparable to, or exceed, that of $r_0$. This suggests that, within the
exponential halo model, a sufficiently long and well-resolved EMRI waveform
could in principle help disentangle the halo's total mass from its radial
extent, offering a strong-field probe of both the amount and the distribution
of dark matter surrounding a supermassive black hole.

\section{Summary and Conclusions}
We have studied the strong-field orbital dynamics and gravitational-wave emission of extreme-mass-ratio inspirals around a Schwarzschild-like black hole dressed by an exponential dark matter halo. Although the halo mass $M_0$ and scale radius $r_0$ enter the metric through the same effective mass function $A(r)$, they turn out to reshape the geodesic structure in genuinely different ways. In the bound-orbit $(E,L)$ phase space, increasing $M_0$ simply translates the allowed region toward larger $L$ with little change in its width, whereas increasing $r_0$ shifts it toward smaller $L$, visibly narrows it, and raises the minimum energy set by the ISCO. The two parameters also pull the periodic orbits in opposite directions: at fixed angular momentum, a larger $M_0$ compresses the orbit while a larger $r_0$ enlarges it; at fixed energy, the two roles reverse.

This distinction carries over directly to the emitted gravitational waves. Within the numerical kludge framework, $r_0$ is clearly the dominant parameter for the waveform, stretching the radial period and driving a progressive dephasing of successive whirl bursts. $M_0$, by contrast, barely changes the waveform amplitude, even at $M_0=0.4$; its signature shows up instead in the timing and phase of the whirl bursts, and only becomes clearly resolved once $M_0$ makes up a substantial fraction of the black hole's mass. Because $r_0$ leaves its mark on both the amplitude and the phase of the signal while $M_0$ affects mainly the phase, a sufficiently long, well-resolved EMRI observation could in principle separate the halo's total mass from how it is spatially distributed.

Taken together, these results support the idea that EMRIs are genuine probes of the dark matter environment around a supermassive black hole, not only its presence but, potentially, its mass and spatial extent separately. Extending this analysis to a rotating background, including radiation reaction in the inspiral, and estimating how well $M_0$ and $r_0$ could actually be measured from a LISA-band signal are natural next steps.

\end{document}